\documentclass[]{aastex631}

\shorttitle{Atmospheric Escape on the TRAPPIST-1 Planets}
\shortauthors{Gialluca et al.}
\graphicspath{{./}{figures/}}
\usepackage{amsmath,amssymb}
\usepackage{wrapfig}
\usepackage{tabularx}
\usepackage{booktabs}
\usepackage[version=4]{mhchem} % Molecular species! e.g. \ce{H2O}

\newcommand\ddfrac[2]{\cfrac{\displaystyle #1}{\displaystyle #2}}
\newcommand{\PeerReview}[1]{{#1}}

\begin{document}

\title{The Implications of Thermal Hydrodynamic Atmospheric Escape on the TRAPPIST-1 Planets}

\author[0000-0002-2587-0841]{Megan T. Gialluca}
\affiliation{Department of Astronomy and Astrobiology Program, University of Washington, Box 351580, Seattle, Washington 98195, USA}
\affiliation{NExSS Virtual Planetary Laboratory, Box 351580, University of Washington, Seattle, Washington 98195, USA}

\author{Rory Barnes}
\affiliation{Department of Astronomy and Astrobiology Program, University of Washington, Box 351580, Seattle, Washington 98195, USA}
\affiliation{NExSS Virtual Planetary Laboratory, Box 351580, University of Washington, Seattle, Washington 98195, USA}

\author[0000-0002-1386-1710]{Victoria S. Meadows}
\affiliation{Department of Astronomy and Astrobiology Program, University of Washington, Box 351580, Seattle, Washington 98195, USA}
\affiliation{NExSS Virtual Planetary Laboratory, Box 351580, University of Washington, Seattle, Washington 98195, USA}

\author{Rodolfo Garcia}
\affiliation{Department of Astronomy and Astrobiology Program, University of Washington, Box 351580, Seattle, Washington 98195, USA}
\affiliation{NExSS Virtual Planetary Laboratory, Box 351580, University of Washington, Seattle, Washington 98195, USA}

\author{Jessica Birky}
\affiliation{Department of Astronomy, University of Washington, Box 351580, Seattle, Washington 98195, USA}

\author{Eric Agol}
\affiliation{Department of Astronomy and Astrobiology Program, University of Washington, Box 351580, Seattle, Washington 98195, USA}
\affiliation{NExSS Virtual Planetary Laboratory, Box 351580, University of Washington, Seattle, Washington 98195, USA}



\begin{abstract}

% This one is 249 words, please check for clarity and "mangleness". 
JWST observations of the 7-planet TRAPPIST-1 system will provide an excellent opportunity to test outcomes of stellar-driven evolution of terrestrial planetary atmospheres, including  atmospheric escape, ocean loss and abiotic oxygen production. While most previous studies use a single luminosity evolution for the host star, we incorporate observational uncertainties in stellar mass, luminosity evolution, system age, and planetary parameters to statistically explore the plausible range of planetary atmospheric escape outcomes. We present probabilistic distributions of total water loss and oxygen production as a function of initial water content, for planets with initially pure water atmospheres and no interior-atmosphere exchange. We find that the interior planets are desiccated for initial water contents below 50 Earth oceans. For TRAPPIST-1e, f, g, and h, we report maximum water loss ranges of 8.0$^{+1.3}_{-0.9}$, 4.8$^{+0.6}_{-0.4}$, 3.4$^{+0.3}_{-0.3}$, and 0.8$^{+0.2}_{-0.1}$ Earth oceans, respectively, with corresponding maximum oxygen retention of 1290$^{+75}_{-75}$, 800$^{+40}_{-40}$, 560$^{+30}_{-25}$, and 90$^{+10}_{-10}$ bars. We explore statistical constraints on initial water content imposed by current water content, which could inform evolutionary history and planet formation. If TRAPPIST-1b is airless while TRAPPIST-1c possesses a tenuous oxygen atmosphere, as initial JWST observations suggest, then our models predict an initial surface water content of \PeerReview{8.2$^{+1.5}_{-1.0}$ Earth oceans} for these worlds, leading to the outer planets retaining \PeerReview{$>$1.5 Earth oceans} after entering the habitable zone.  Even if  TRAPPIST-1c is airless, surface water on the outer planets would not be precluded.

\end{abstract}

%\keywords{)}


\section{Introduction} \label{sec:intro}
% TOPIC:  M dwarf planets important for questions of life's distribution, and observationally accessible in the near-term.
As M dwarfs are the most common stars in our local stellar neighborhood \citep{Cantrell2013mstarprevalence}, whether their planetary systems can harbor life is a key question in astrobiology that may be amenable to observational tests in the near term. Due to their small size and low temperature, it is considerably easier to conduct observations of terrestrial planets in M dwarf systems with current and near-term technology than for any other main sequence host star \citep{tarter2007reappraisal}. Terrestrial planetary targets of interest for atmospheric characterization with M dwarf hosts may be accessible with the JWST \citep{gardner2006james,seager2009transiting,deming2009discovery,Morley2017jwstdetect,lustig2019detectability,wunderlich2019detectability,gialluca2021characterizing}, and the ground-based extremely large telescopes (ELTs; e.g., E-ELT, TMT, GMT) \citep{gilmozzi2007european,Snellen2013groundbased,marconi2016eelt,Ezaki2016TMToverview,Fanson2018GMToverview,LopezMorales2019eltdetect,Currie2023ELT,HardegreeUllman2023bioverse}.  In the more distant future, the flagship Habitable Worlds Observatory (HWO), as defined by the recent Astronomy 2020 Decadal Survey, may include a small subset of M dwarf planets in its observing sample, that will be accessible via transmission and/or direct imaging \citep{luvoir2019luvoir,gaudi2020habitable,national2021decadal}.

%TOPIC:  Despite observational advantages, have tons of habitability challenges due to the star. 
Despite the observational advantages of M dwarf systems, they face a number of habitability challenges due to the proximity of the habitable zone (HZ) to the parent star. This leads to strong gravitational interactions, such as tidal locking, that can result in synchronous rotation and extreme day-night temperature gradients \citep{Showman2013circulationreview,Yang2014tidallockwatertrap,Barnes2017tidallock,Lobo2023limitMwater}.  The M dwarf star's relative proximity to its HZ planets also enhances radiative interactions, and these could pose the most significant habitability and characterization challenges.  The early and pronounced luminosity evolution of the star may produce a runaway greenhouse on the planets \citep{Ingersoll1969runaway,Pollack1971runawayvenus,Nakajima1992runawaystudy,Goldblatt2013runawaylimit,Barnes2013tidalvenuses,Turbet2019runawayradiusinflation}, potentially promoting an extended magma ocean phase \citep{Hamano2013typesmagmaoc,schaefer2016predictions,barth2021magma} and severe atmospheric mass loss through thermal atmospheric escape \citep[e.g.,][]{hunten1973escape}.  In addition to luminosity evolution, heightened stellar activity also increases the stellar XUV of M dwarf stars, which enhances atmospheric loss, and may complicate characterization and interpretation of spectra from these planets by driving planetary photochemistry that can enhance, destroy or even mimic proposed biosignatures \citep{Segura2005Mstarbiosignatures,meadows2018o2InContext}.

% TOPIC: PMS phase is the most intense period of luminosity evolution and will affect habitability of HZ planets
During the extended pre-main sequence (PMS) phase, an M dwarf is at its brightest as it contracts to its main sequence size \citep{Baraffe2015stellartracks}.
%,  The most intense star-planet interactions occur due to an M dwarf's heightened XUV environment during the pre-main sequence phase \citep{LugerBarnes2015extreme}. During the PMS phase,
The protostar reaches radiative equilibrium but not nuclear fusion, leading to extended Kelvin-Helmholtz contractions to compensate for the loss of energy through radiation \citep{Hayashi1961stellarevolve,LamersLevesqueStellar2017}. These contractions lead to a decrease in radiative opacity, allowing energy to escape more rapidly while the star continues to accrete mass from the surrounding disk -- both effects contribute to the star being orders of magnitude brighter than on the main sequence \citep[e.g.,][]{Baraffe1998,dotter2008dartmouth,reid2013stars,LamersLevesqueStellar2017}. Additionally, the PMS phase is considerably longer for less massive stars; while a Sun-like star may spend tens of millions of years in the PMS, an M dwarf may spend on the order of a billion years in the PMS \citep{Baraffe2015stellartracks}.

% TOPIC: introducing **hydrodynamic** thermal escape and noting the effects it may have for a planet (that will be in the HZ on the MS) - water loss and oxygen production - and why those things are challenging for characterization (e.g., false positive)

This increased luminosity and XUV activity during the PMS phase of an M dwarf results in prolonged incident radiation that is well in excess of the runaway greenhouse limit on planets that will be in the HZ during the main sequence. This early PMS phase may drive thermal hydrodynamic atmospheric escape, which can strongly abrade a planetary atmosphere, and modify the redox state \citep{Jeans1925Escape,hunten1973escape,lammer2003atmospheric,Volkov2011thermalescapetransition,LugerBarnes2015extreme}. In the hydrodynamic  escape regime, incident radiation leads to water photolysis in the atmosphere and the hydrogen is subsequently lost, leading to permanent water loss and the potential for abiotic oxygen build-up  \citep{Wordsworth2014oxygenDomAtmHZ,LugerBarnes2015extreme,lincowski2018Trappist}.  Since oxygen production from ocean-loss could produce thousands of bars of \ce{O2} \citep{LugerBarnes2015extreme,lincowski2018Trappist}, with at least a few bars retained after atmospheric, crustal, and magma ocean losses \citep{schaefer2016predictions,barth2021magma}, hydrodynamic escape on planets orbiting M dwarfs could produce significant abiotic oxygen that is potentially detectable and can be confused for biologically-generated  \ce{O2} \citep[e.g.,][]{segura2007abiotic,schwieterman2016identifying,Schwieterman2016erratum,meadows2017reflections,meadows2018o2InContext}. 
%Moreover, an M dwarf's luminosity evolution may leave post-ocean-loss planets with high abiotic \ce{O2} atmospheres in the habitable zone \citep{lincowski2018Trappist}, where they will be harder to distinguish from habitable planets with biospheres.
Consequently, it is highly important to understand the mechanisms that generate abiotic \ce{O2}, and to quantify the likely range of oxygen production that may occur during this hydrodynamic escape process.
 
The 7-planet TRAPPIST-1 system \citep{gillon2016temperate,Gillon2017sevenplanets} provides an excellent, nearby (12 pc), and relatively observationally accessible  example of a transiting  M dwarf system with which we can explore and test the outcomes of enhanced atmospheric escape for terrestrial-sized planets.   As a late-type M8V star, TRAPPIST-1 likely provided an early heightened XUV environment \citep{Baraffe2015stellartracks}, and its 7 transiting terrestrial planets, including 3 -- 4 within the current habitable zone, enable tests of atmospheric loss processes as a function of distance from the parent star.  As several of these planets may follow similar evolutionary pathways as Venus or Earth, there is also a prime opportunity for comparative planetology, particularly in questioning what makes a potentially habitable planet inhospitable \citep{Kane2019venuslaboratory}.  

TRAPPIST-1 is a high priority target for JWST General and Guaranteed Time Observations \citep{Gillon2020trappistjwstcommunityinit,Greene2023T1bJWST,Zieba2023T1cJWST}, and quantifying likely water loss ranges and abiotic oxygenation histories will be needed to interpret current and upcoming observations. 
%while also representing a high priority target for the community due to its close proximity and the existence of 7 transiting terrestrial plantets, 3 -- 4 of which may be in the habitable zone. 
While possible post-ocean and atmospheric loss atmospheres, and habitable environments in this system have already been modeled in several previous studies \citep{lincowski2018Trappist,krissansen2018detectability,lustig2019detectability,barth2021magma}, until recently \citep{birky2021improved,krissansen2022predictions} a lack of constraints on the XUV evolution have prohibited comprehensive statistical studies of PMS evolution outcomes.

Whereas previous studies \citep[e.g., ][]{LugerBarnes2015extreme,tian2015atmospheric,schaefer2016predictions,bolmont2017water,bourrier2017temporal,lincowski2018Trappist,Johnstone2020hydrodynamicesc} have explored implications for planetary evolution by analyzing planetary escape rates driven by single or tens of stellar luminosity evolution tracks, 
%to better understand the theoretical implications of this form of evolution, 
more comprehensive statistical studies are needed to constrain the broader family of evolutionary solutions that are consistent with preexisting data. For example, while a single stellar evolutionary track may provide a plausible history for water loss on a planet, it cannot be used to explore the phase space of water loss and oxygen production outcomes, quantify confidence in the single interpretation, or place upper limits on maximum water loss and abiotic \ce{O2} production. 
By leveraging the computationally efficient planetary evolution model, VPLanet \citep{Barnes2020VPLanet}, and developing support scripts (Vconverge) to automatically run simulation suites to convergence, here we explore the full observational uncertainty of the stellar evolution \citep{birky2021improved} and planetary parameters \citep{agol2021refining} of the TRAPPIST-1 system. To define the plausible phase space of water loss and oxygen production outcomes, we conduct hundreds of thousands of simulations of the water evolution history of the TRAPPIST-1 planets across a grid of initial water contents from 1 terrestrial ocean (TO) to 500 TO - spanning the full range that has been suggested as plausible in the literature \citep{Schoonenberg2019t1formation,agol2021refining,Acuna2021T1Hydrospheres,Raymond2022upperlimT1}. A terrestrial ocean, or Earth ocean, is our standard unit of measurement for water content and is equivalent to 1.39 $\times$ 10$^{21}$ kg. \PeerReview{Though \citet{krissansen2022predictions} adopted a similar statistical approach considering the system constraints from \citet{birky2021improved} and \citet{agol2021refining}, our study conducts a greater number of simulations to more comprehensively explore the phase space of initial water content, although at the expense of a coupled atmosphere and interior}. We present predictions with 1$\sigma$ uncertainty on water loss as a function of initial water abundance for each TRAPPIST-1 planet. This study additionally identifies the maximum water loss possible for the HZ planets regardless of increasing initial water mass, assuming initially pure water atmospheres and no exchange with the interior. Furthermore, given this maximum water loss and our understanding of oxygen drag during escape, we calculate the maximum amount of abiotic oxygen that could be produced through this process, with perfectly ineffective oxygen sinks at the surface, to provide upper limits on total atmospheric oxygen build-up. 

%%% Saving old text:
%%%%% THE SENTIMENT HERE MAY BE MISSING NOW AND MAY NEED TO BE WORKED BACK IN
% Statistical approaches to planetary evolution modeling have become more of a focus in the field \citep[e.g.,][]{krissansen2022predictions}, but our study differs by exploring initial water content on a broad grid as opposed to a random sampling method; this choice additionally means we run similar numbers of simulations for each initial water content that other studies may run total.

%Here is all of the wonderful science that is enabled!  This is now the specifics of what a statistical study can get us (complements generic arguments). 
%%% TOPIC: okay we've made these distributions, heres all the insight we can now gain from this statistical study that will be presented in the paper.
% Needs to include (even briefly) (not in particular order here): comparative planetology of T1 planet outcomes as a function of orbital distance, method for retrieving evolutionary history (discussion 4.4.1 uses phrase "infer system evolution from observation" or "retrieve evolutionary history", use this language to be consistent), interpreting JWST results (which can be single planet or strengthened through comparative planetology, I think the latter is important to highlight given our strategy of future JWST proposals), model intercomparison
Through this statistical study of all 7 known planets in the system, escape evolution as a function of orbital distance is explored, methods of inferring system evolution from observation are developed and demonstrated, model intercomparisons are explicitly quantified \citep{lincowski2018Trappist,krissansen2022predictions}, and we begin to interpret the most recent JWST observations \citep{Greene2023T1bJWST,Zieba2023T1cJWST} in the context of our findings. To infer system evolution, we show how our knowledge of water loss over time can be used in conjunction with present-day water mass fraction (WMF) measurements of the TRAPPIST-1 planets to retrieve plausible evolutionary histories and probable initial water contents. The retrieval of evolutionary history will aid in developing a relationship between initial and final water content that can connect formation models to observations. In regards to JWST, we consider recent observations \citep{Greene2023T1bJWST,Zieba2023T1cJWST} in the context of our results by analyzing the planets individually, and then by showing the significant additional information that can be discerned when considering multiple bodies in the system and using a comparative planetology approach. In particular, we provide insights into initial water content of the system and current plausible conditions on planets that have yet to be observed.

Section \ref{sec:methods} describes our modeling methods, including a description of the atmospheric escape mechanisms we consider (\S \ref{subsec:model}), construction of model input (\S \ref{subsec:Input}), how convergence is achieved through multiple simulations (\S \ref{subsec:Convergence}), and a comparison to the model before these most recent modifications \citep[as presented in ][]{lincowski2018Trappist} (\S \ref{subsec:ModelCompare}). Section \ref{sec:Results} presents the key results of our study, including the evolutionary pathways observed in our simulations (\S \ref{subsec:EscapeTime}), the full statistical spread of all models (\S \ref{subsec:Statistical}), identification of the most probable initial water content to reproduce final water mass fractions presented in \citet{agol2021refining} (\S \ref{subsec:InitialWaterPredict}), quantification of oxygen retention efficiency (\S\ref{subsec:O2Efficiency}), and a comparison of our results to a similar study published in \citet{krissansen2022predictions} (\S\ref{subsec:JoshKTCompare}). The interpretation, significance, and limitations of these results and comparison to past work is discussed in Section \ref{sec:Discussion}, notably including an analysis of recent JWST observations in the context of our study (\S\ref{subsubsec:JWSTObsDiscuss}). We summarize our conclusions in Section \ref{sec:Conclusion}. Please note, throughout the paper, in referring to individual planets, we adopt the shorthand of `T1-$i$' for `TRAPPIST-1$i$', where $i$ would refer to the planet in question (b, c, d, e, f, g, or h). 

\section{Methods} \label{sec:methods}

We conduct our simulations using VPLanet%\PeerReview{\footnote{https://github.com/VirtualPlanetaryLaboratory/vplanet}}
, an open source code for modeling star, planet, and planetary system interactions and their impact on evolution \citep{Barnes2020VPLanet}. VPLanet is broken up into modules for different physical processes; in this work, the STELLAR (stellar evolution, \S \ref{subsubsec:stellarmethods}) and AtmEsc (atmospheric escape, \S \ref{subsubsec:atmescmethods}) modules were used, and we specifically modify the treatment of diffusion-limited escape, the critical drag XUV flux, and the allowed behavior of the ratio of oxygen to hydrogen loss. A mathematically detailed description of our modifications and how they operate in AtmEsc can be found in Appendix \ref{appendix:atmesc}. 

Input planetary and stellar parameters for our models come from \citet{agol2021refining} and \citet{birky2021improved}, respectively. Using these updated constraints, we build and preserve self-consistent relationships between posteriors (see \S \ref{subsec:Input}) allowing us to randomly sample a string of self-consistent planetary and stellar inputs for each simulation, as opposed to each input individually. Typically, 3500 -- 5500 simulations are conducted per initial water content per planet to build distributions of final water and oxygen content. The exact number of simulations in a particular suite is determined by model convergence (found via the Kolmogorov–Smirnov, or K-S, test), which is described in Section \ref{subsec:Convergence}. In Appendix \ref{subsec:ModelCompare}, we illustrate the effects of our modifications to AtmEsc by comparing it to a simulation run with an older, validated version of the model \citep{LugerBarnes2015extreme,lincowski2018Trappist,Lincowski2022eratum}.

We note that a small number of outliers stemming from $\sim$1.24\% of the stellar mass samples were identified and removed from our distributions. These outlying stellar mass samples were greater than 4 standard deviations from the average and were likely a product of the machine learning method implemented in \citet{birky2021improved}. A more detailed description of outlier removal is given in Appendix \ref{appendix:outliers}.

%Note also, a small number of outliers were identified and removed from our final dataset, the description and justification of this removal can be found in Section \ref{subsec:outliers}. Finally, we describe in Section \ref{subsec:initialpredictionMethods} a method for identifying the most probable initial water content predicted by our model given a known present-day state, which is implemented in the results of Section \ref{subsec:InitialWaterPredict}. 

\subsection{Model Descriptions} \label{subsec:model}

\subsubsection{Stellar Evolution (STELLAR)} \label{subsubsec:stellarmethods}

%The STELLAR module evolves the host star environment and physical parameters over time. 

The STELLAR module evolves the host star characteristics over time by performing a bicubic spline integration across mass and time of the evolutionary tracks in \citet{Baraffe2015stellartracks} to track the stellar radius, radius of gyration, effective temperature, and luminosity from the PMS phase to present-day. It then uses the empirical broken power law formula of \citet{Ribas2005evolution} to track the XUV output as a function of bolometric luminosity, which is particularly relevant for M dwarfs such as TRAPPIST-1:
\begin{equation}
    \large
    \ddfrac{L_{XUV}}{L_{Bol}} = 
    \begin{cases}
        f_{sat} & \text{if } t \leq t_{sat} , \\
        f_{sat} \left(\ddfrac{t}{t_{sat}}\right)^{-\beta_{XUV}} & \text{if } t > t_{sat} ,
    \end{cases}
\end{equation}
where $L_{XUV}$ is the XUV luminosity, $L_{Bol}$ is the bolometric luminosity, $t$ is the time, $t_{sat}$ is the XUV saturation time, $\beta_{XUV}$ is the exponential decay rate of the XUV luminosity, and $f_{sat}$ is the XUV saturation fraction, which is the ratio of XUV to bolometric luminosity during the saturation time; the latter 3 of these parameters are user defined inputs (see \S \ref{subsec:Input}). The STELLAR module was used to fit the TRAPPIST-1 stellar input parameters that we use in this study \citep{Fleming2020stellarevolution,birky2021improved}, and a more thorough and mathematical description of STELLAR can be found in \citet{Barnes2020VPLanet}; STELLAR was not modified in this work. The output stellar parameters from the STELLAR module become the input for the atmospheric escape model, AtmEsc, at each timestep; the stellar XUV characteristics, evolved by STELLAR, are the main driver of hydrodynamic thermal atmospheric escape. 

\subsubsection{Atmospheric Escape (AtmEsc)} \label{subsubsec:atmescmethods}

Atmospheric escape can typically be split up into the thermal and non-thermal regimes; due to the heightened XUV environment, the TRAPPIST-1 planets are most likely subjected to primarily thermal escape processes during the PMS phase. \PeerReview{For the innermost planets that never enter the habitable zone, thermal escape dominates throughout the lifetime of the atmosphere, but for the remaining planets, once the star reduces in luminosity sufficiently for them to enter the habitable zone, non-thermal escape processes have the potential to be more significant \citep[e.g.,][]{Dong2018atmescT1}}. Within the thermal regime we can consider Jeans escape or hydrodynamic blow-off -- the latter occurring when the internal energy of the escaping species approaches the kinetic energy required for escape \citep{hunten1973escape}; hydrodynamic loss is likely to provide a larger impact on the TRAPPIST-1 planetary atmospheres than Jeans escape, due to its extremely high mass loss rates. 

Hydrodynamic escape can further be categorized as energy- or diffusion-limited. Energy-limited escape occurs when the escape flux is limited by the amount of energy incident on the planet, whereas diffusion-limited escape occurs when the escape flux is limited by the speed at which the escaping gas (typically hydrogen) can diffuse through a static background gas. During energy-limited escape, the escaping species may also drag a heavier species with it -- the efficiency of which is highly dependent on mixing ratios in the atmosphere, the incoming XUV flux, and the mass of the species being dragged; thus, it can lead to observable effects such as mass fractionation of the heavier dragged species \citep[e.g.,][]{Hunten1987massfracinescape,Zahnle1990massfracinescape}, but this process produces only very weak fractionation of the hydrodynamically escaping species (i.e., D/H ratio, $^{18}$O/$^{16}$O) \citep{Mandt2009isotopicFract,Lammer2020FractionationReview}. This drag can also be a considerable source of atmospheric mass loss, particularly during a star's most active period. In the diffusion-limited regime, the escape flux must be less than the maximum upward flux for which the background atmosphere is static \citep{hunten1973escape}; thus, by definition, oxygen drag cannot occur under diffusion-limited escape. In the diffusion-limited regime mass fractionation may occur \citep{Hunten1989escapereviewpaper,Zahnle1990massfracinescape}, enhancing the atmosphere's D/H ratio, but is likely only considerable on large timescales \citep{Mandt2009isotopicFract}.

Here we use VPLanet's AtmEsc module  to model both energy- and diffusion-limited escape for the TRAPPIST-1 planets, and a full description of this module can be found in \citet{Barnes2020VPLanet}, with a more succinct mathematical description of its implementation in the context of this work provided in Appendix \ref{appendix:atmesc}. In comparison to previous work with AtmEsc \citep{LugerBarnes2015extreme,lincowski2018Trappist,Lincowski2022eratum,Barnes2020VPLanet}, we modify the treatment of the switch to diffusion-limited escape, implement a calculation of the critical drag XUV flux (Eq. \ref{eq:CritDrag}), and change the allowed behavior of $\eta$ (ratio of oxygen escape to production, Eq. \ref{eq:etao}) to be less restrictive; these changes follow \citet{schaefer2016predictions}. In this work, the switch from energy- to diffusion-limited escape occurs when the mixing ratio of free oxygen (i.e., oxygen not contained in water molecules) is greater than or equal to the mixing ratio of water. The critical drag XUV flux ($F_{XUV}^{Crit}$, Eq. \ref{eq:CritDrag}) is defined as the minimum incoming XUV flux required for oxygen drag to occur; if the incoming XUV flux is less than this, oxygen drag will be halted regardless of whether or not the escape is energy-limited. Finally, $\eta$ is defined as the ratio of oxygen escape to production, or the ratio of oxygen to hydrogen loss. In past AtmEsc implementations, this was restricted to be $\le$1, meaning, at most, 1 oxygen atom could be lost for every 2 hydrogen atoms. Here, we allow $\eta$ to vary up to $\le$2 as we note that an accumulated oxygen reservoir may be depleted under highly favorable drag conditions, effectively meaning oxygen is lost faster than it is produced for certain periods of time. \PeerReview{Appendix \ref{subsec:ModelCompare} shows a comparison of the modified model used in this work to the previous version of the model \citep{LugerBarnes2015extreme,lincowski2018Trappist,Lincowski2022eratum}; in general, our modifications lead to slightly higher water loss ($\sim$1 -- 8\%) and thus slight higher oxygen build-up ($\sim$10\%)}. As simplifications are made within this implementation, Section \ref{subsubsec:modeldiscuss} discusses future model improvements that could be made for a more realistic escape treatment. 

% suggest putting this paragraph below the following one. ---> Did this!
Our simulations make several assumptions, including initially pure water atmospheres, no surface oxygen sinks or geological activity, and that all water is on the planetary surface and available for escape by 1 Myr. \PeerReview{As in \citet{Wordsworth2018redox}, we assume no primordial H$_{2}$-He envelope to be conservative in finding upper limit values of water loss and oxygen production.} We explore initial water contents from 1 terrestrial ocean (TO) to 500 TO on a modified log space grid. \PeerReview{We choose the upper limit for initial water content simulated here to be 500 TO, based on several recent studies which constrain the initial and present-day water mass fractions and suggest these are around 10\%wt \citep[e.g.,][]{Schoonenberg2019t1formation,agol2021refining,Raymond2022upperlimT1}, which would be equivalent to $\sim$430 TO on a 1 Mearth planet; 500 TO is approximately bracketed between 10\%wt of the larger planets, T1-b, c, and g, and the smaller planets, T1-d, e, f, and h.} 
%A terrestrial ocean, or Earth ocean, is our standard unit of measurement for water content and is equivalent to 1.39 $\times$ 10$^{21}$ kg. 
As we assume all oxygen not lost through escape remains in the atmosphere, our standard unit of measurement for oxygen production is bars of atmospheric pressure, which considers the mass and radius of a particular planet when converting from kg of oxygen. We additionally assume uniform mixing ratios for gases in the atmosphere (no vertical structure) with an isothermal temperature profile and do not consider photochemical processes besides the division of water molecules through photolysis. The effects our assumptions may have on the results is explored in Section \ref{subsec:LimitationsDiscuss}.

\subsection{Input Description} \label{subsec:Input}

In determining simulation inputs, we consider not only observational uncertainty in the stellar luminosity evolution and physical planetary parameters \citep{birky2021improved,agol2021refining}, but also that planetary parameters such as mass, radius, orbit and insolation can be self-consistently dependent on the assumed stellar environment, as planetary mass and radius are derived from planet-to-star ratios. To account for this, we construct self-consistent chains of input parameters for the entire system, such that one simulation randomly samples an input chain as opposed to each individual parameter separately. 
%To fully account for the observational uncertainty in the stellar and planetary parameters \citep{birky2021improved,agol2021refining}, and because we need to consider that planetary parameters such as mass, radius, orbit and insolation can be self-consistently dependent on stellar parameters given that planetary mass and radius are derived from planet to star ratios, we construct chains of input parameters for the entire system; in this way, one simulation randomly samples a self-consistent input chain as opposed to each individual parameter separately. In defining an input string, we randomize sampling from distributions of each individual input, which are then combined to preserve a self-consistent, meaningful relationship.

Here we detail the stellar and planetary parameters we consider, and how the input chain construction for a single simulation is accomplished, which is then repeated for simulation suites at a particular initial water content.
%This process is facilitated by Vspace, an open source support routine for VPLanet \citep{Barnes2020VPLanet} that creates a suite of input files for the model across a parameter space defined by the user. The process subsequently described in this section is new functionality that has been added to Vspace in this work to allow the program to randomly sample from a distribution of potential inputs. 
The stellar inputs used in our work \citep[from][]{birky2021improved} are the mass of the star ($M_{*}$), the XUV saturation fraction ($f_{sat}$), the XUV saturation time ($t_{sat}$), the exponential decay rate of the XUV luminosity ($\beta_{XUV}$), and the age ($\tau$). For each of the 7 planets, the input parameters used \citep[from][]{agol2021refining} are the planetary mass ratio ($M_{p}/M_{*}$), the planetary radius ratio ($R_{p}/R_{*}$), the eccentricity ($e$), and the orbital period ($P$). More specifically, \citet{agol2021refining} presents a transit timing and photodynamical analysis of the system; the former, transit timing variations (TTV), are highly sensitive to the ratios of planetary to stellar mass, along with the eccentricity and orbital period. The photodynamical analysis provides sensitive constraints on the stellar density ($\rho_{*}$) and planet-to-star radius ratios by combining the mass ratios and orbital parameters derived from the TTV analysis to improve the derivation of planetary and stellar densities from Spitzer photometry \citep{agol2021refining}. Thus, the actual planetary mass and radius of a single input chain will be dependent on the sampled stellar parameters.

To construct one input chain, one suite of stellar, TTV, and photodynamical inputs are randomly sampled. The selected stellar mass is then combined with the stellar-to-planetary mass ratios from the TTV sample to produce the planetary masses:
\begin{equation}
\label{planetmass}
    \large
    M_{p,i} = \frac{M_{p,i}}{M_{*}}M_{*} \ ,
\end{equation}
where $i$ refers to the planet in question (b, c, d, e, f, g, or h). Using Eq. (\ref{planetmass}) to find the planetary mass creates a relationship between the stellar inputs and the TTV inputs. The same stellar mass is then used to find the stellar radius using the stellar density from the photodynamical inputs \citep{agol2021refining}:
\begin{equation}
\label{stellarradius}
    \large
    R_{*} = \left( \frac{3 M_{*}}{4 \pi \rho_{*}} \right)^{\frac{1}{3}} \ .
\end{equation} 
Finally we can take this stellar radius and combine it with our stellar-to-planetary radii ratios to get the planetary radii:
\begin{equation}
\label{planetaryradius}
    \large
    R_{p,i} = \frac{R_{p,i}}{R_{*}}R_{*} \ .
\end{equation}

Thus, by using one sampled stellar mass and Equations (\ref{planetmass}) through (\ref{planetaryradius}), a self-consistent relationship between all priors is built. For our constructed set of input chains, no single stellar, TTV, or photodynamical sample from \citet{birky2021improved} and \citet{agol2021refining} is used more than once; and likewise, no one input chain is used more than once in a given simulation suite (i.e., at a given initial water content). The process of sampling input chains is facilitated by VSPACE \citep{Barnes2020VPLanet}, an open source support routine for VPLanet that creates a suite of input files for the model across a parameter space defined by the user; sampling from user-defined input distributions is new functionality that has been added to VSPACE in this work. Finally, Tables \ref{tab:planetparams} and \ref{tab:starparams} show the 50$^{th}$ percentile values and 1-$\sigma$ uncertainties of the system's physical parameters, found from 20,900 input chains using the probability distributions described in \citet{agol2021refining} and \citet{birky2021improved}. Since we fold in uncertainty in the stellar mass \citep[as reported by][]{birky2021improved}, our planetary mass and radii uncertainties reported in Table \ref{tab:planetparams} may differ slightly than those reported in \citet{agol2021refining}, despite using identical mass and radius ratios, because they adopted a fixed 0.09 M$_{\bigodot}$.

\begin{table}[h]
    \centering
    \hspace*{-2cm}\begin{tabular}{c|c|c|c|c}
        Body & Mass [$M_{\bigoplus}$] &  Radius [$R_{\bigoplus}$] & Eccentricity & Orbital Period [Days] \\
        \toprule
        T1-b & 1.377$_{-0.062}^{+0.063}$ & 1.117$_{-0.007}^{+0.009}$ & 0.00401$_{-0.00233}^{+0.00266}$ & 1.510826$_{-0.000005}^{+0.000005}$ \\
        \midrule
        T1-c & 1.311$_{-0.045}^{+0.044}$ & 1.098$_{-0.008}^{+0.010}$ & 0.00190$_{-0.00133}^{+0.00192}$ & 2.421936$_{-0.000017}^{+0.000017}$ \\
        \midrule
        T1-d & 0.388$_{-0.008}^{+0.008}$ & 0.789$_{-0.008}^{+0.009}$ & 0.00583$_{-0.00141}^{+0.00129}$ & 4.049219$_{-0.000025}^{+0.000026}$ \\
        \midrule
        T1-e & 0.693$_{-0.014}^{+0.013}$ & 0.922$_{-0.009}^{+0.010}$ & 0.00637$_{-0.00080}^{+0.00103}$ & 6.101013$_{-0.000034}^{+0.000035}$ \\
        \midrule
        T1-f & 1.040$_{-0.017}^{+0.017}$ & 1.046$_{-0.009}^{+0.010}$ & 0.00848$_{-0.00118}^{+0.00103}$ & 9.207540$_{-0.000029}^{+0.000030}$ \\
        \midrule
        T1-g & 1.322$_{-0.019}^{+0.019}$ & 1.131$_{-0.009}^{+0.011}$ & 0.00404$_{-0.00091}^{+0.00101}$ & 12.352445$_{-0.000052}^{+0.000054}$ \\
        \midrule
        T1-h & 0.325$_{-0.018}^{+0.019}$ & 0.756$_{-0.012}^{+0.012}$ & 0.00370$_{-0.00072}^{+0.00068}$ & 18.772868$_{-0.000209}^{+0.000217}$ \\
    \end{tabular}
    \vspace{2mm}
    \caption{50$^{th}$ percentile values for all planetary parameters used as input found from 20900 samples. Uncertainties given are set by 16$^{th}$ and 84$^{th}$ percentile values (1$\sigma$). Values come from transit timing and photodynamical analysis in \citet{agol2021refining} combined with stellar mass from \citet{birky2021improved}.}
    \label{tab:planetparams}
\end{table}

%\begin{table}[h]
%    \centering
%    \begin{tabular}{c|c|c|c|c|c|c|c}
%        Body    &   T1-b    &   T1-c    &   T1-d    &   T1-e   &   T1-f  &   T1-g   &   T1-h \\
%        \toprule
%        Mass [$M_{\bigoplus}$] & 1.377$_{-0.062}^{+0.063}$ & 1.311$_{-0.045}^{+0.044}$ & 0.388$_{-0.008}^{+0.008}$ & 0.693$_{-0.014}^{+0.013}$ & 1.040$_{-0.017}^{+0.017}$ & 1.322$_{-0.019}^{+0.019}$ & 0.325$_{-0.018}^{+0.019}$  \\
%        \midrule
%        Radius [$R_{\bigoplus}$] & 1.117$_{-0.007}^{+0.009}$ & 1.098$_{-0.008}^{+0.010}$ & 0.789$_{-0.008}^{+0.009}$ & 0.922$_{-0.009}^{+0.010}$ & 1.046$_{-0.009}^{+0.010}$ & 1.131$_{-0.009}^{+0.011}$ & 0.756$_{-0.012}^{+0.012}$ \\
%        \midrule
%        Eccentricity & 0.00401$_{-0.00233}^{+0.00266}$ & 0.00190$_{-0.00133}^{+0.00192}$ & 0.00583$_{-0.00141}^{+0.00129}$ & 0.00637$_{-0.00080}^{+0.00103}$ & 0.00848$_{-0.00118}^{+0.00103}$ & 0.00404$_{-0.00091}^{+0.00101}$ & 0.00370$_{-0.00072}^{+0.00068}$ \\
%        \midrule
%        Orbital Period [Days] & 1.510826$_{-0.000005}^{+0.000005}$ & 2.421936$_{-0.000017}^{+0.000017}$ & 4.049219$_{-0.000025}^{+0.000026}$ & 6.101013$_{-0.000034}^{+0.000035}$ & 9.207540$_{-0.000029}^{+0.000030}$ & 12.352445$_{-0.000052}^{+0.000054}$ & 18.772868$_{-0.000209}^{+0.000217}$ \\
%        \bottomrule
%    \end{tabular}
%    \caption{Caption}
%    \label{tab:my_label}
%\end{table}

\begin{table}[h]
    \centering
    \hspace*{-1cm}\begin{tabular}{c|c}
        Parameter & Value \\
        \toprule
        Mass [M$_{\bigodot}$] & 0.090$_{-0.001}^{+0.001}$ \\
        \midrule
        XUV Saturation Fraction [log$_{10}$($L_{XUV}/L_{Bol}$)] & -3.030$_{-0.231}^{+0.251}$ \\
        \midrule
        XUV Saturation Time [Gyr] & 3.146$_{-1.459}^{+2.217}$ \\
        \midrule
        XUV Decay Rate & -1.172$_{-0.277}^{+0.270}$ \\
        \midrule
        Age [Gyr] & 7.962$^{+1.783}_{-1.867}$ \\
    \end{tabular}
    \vspace{2mm}
    \caption{50$^{th}$ percentile values for all stellar parameters used as input found from 20900 samples. Uncertainties given are set by 16$^{th}$ and 84$^{th}$ percentile values (1$\sigma$). Values come from \citet{birky2021improved}.}
    \label{tab:starparams}
\end{table} 

\subsection{Convergence} \label{subsec:Convergence}

%Past studies have looked at the effect of massive thermal atmospheric escape on water retention and oxygen production, often for single or limited samples of stellar evolutionary histories \citep{LugerBarnes2015extreme,tian2015atmospheric,schaefer2016predictions,lincowski2018Trappist,bolmont2017water,bourrier2017temporal,Johnstone2020hydrodynamicesc}, here we greatly expand on these efforts with a  thorough statistical exploration using updated distributions of the TRAPPIST-1 system parameters. This allows us to define plausible ranges of water loss and oxygen production outcomes, and the computational efficiency of VPLanet allows us to explore a wide range of initial water contents.% with little overhead.

To create probability distributions of the final water loss and oxygen production on each planet per initial water content, we systematically add simulation outcomes to a distribution until convergence is achieved, i.e., when the addition of simulations to the sample no longer changes the 3$\sigma$ confidence intervals of the resulting distribution. This process begins with an initial set of 500 simulations, whose final water and oxygen mass comprise the initial distribution of escape outcomes. A step is then taken, defined as the addition of 100 simulation outcomes to the distribution. The difference between the probability distributions from before and after the step is then quantified by the two-sample K-S test, which essentially determines how likely it is that the two distributions originated from the same (unknown) parent distribution. This process of taking steps is continued until the K-S statistic is 0.0004 or less, which we empirically determined to be converged, meaning the addition of more simulations no longer appreciably changes the results. As an added precaution, the model is required to report convergence for 3 consecutive steps before it declares global convergence has been achieved and the result finalized. Typically this process requires approximately 3500 -- 5500 simulations per planet per initial water content. 

This process is facilitated by Vconverge, an open source support routine for VPLanet \citep{Barnes2020VPLanet} that has been created for this work. Vconverge uses the updates made to VSPACE (described in the previous Section \ref{subsec:Input}) to create the input files needed for the initial simulation set, and each subsequent step of 100 simulations. This support routine then automates the process described in this section to find convergence. 

We run simulation suites over a modified log grid of initial water contents spanning 1 -- 500 TO. From 1 -- 20 TO, we use a step size of 1 TO; from 20 -- 100 TO, we use a step size of 10 TO; then from 100 -- 500 TO, we use a step size of 100 TO. In total, 32 initial water contents are considered. In the case of T1-e, an additional 9 simulation suites were run between 100 -- 200 TO, with a step size of 10 TO, to achieve the results shown in Section \ref{subsec:InitialWaterPredict}. These additional simulation suites are only considered with respect to that result, as their inclusion would not affect any other paper result. 

%Figure \ref{fig:Convergence} shows an example of two probability distribution comparisons that our model would consider for T1-e with an initial water content of 20 TO. The left panel of figure \ref{fig:Convergence} shows the distributions from the initial set and the first step representing an unconverged suite while the right panel shows the distributions for the penultimate and final steps representing a converged simulation suite. In this case, the suite shown found global convergence in 3600 simulations.

%\begin{figure}
%    \centering
%    \includegraphics[width=0.9\textwidth]{ConvergencePlot.png}
%    \caption{Two examples of probability distribution comparisons for final water mass of T1-e when given an initial water content of 20 TO. The distributions after the intial set (500 simulations) and the first step (600 simulations total) is shown on the left as an example of an unconverged simulation suite. The right shows the distributions after the penultimate step (3500 simulations total) and the last step (3600 simulations total) as an example of a converged simulation suite.}
%    \label{fig:Convergence}
%\end{figure} 

\section{Results} \label{sec:Results}

%Here we present the results of our simulations, including a description of escape pathways through time, the probabilistic distributions of water loss and oxygen production as a function of initial water content, and several analyses of that data. 
We first describe the possible pathways of escape observed in our simulations for each planet through time, and how they change for different initial water contents (\S \ref{subsec:EscapeTime}). We then present probabilistic distributions of final water loss and oxygen production for each planet as a function of initial water content (\S \ref{subsec:Statistical}). Finally, we show the results of analyses performed with the data, including methodology for predicting initial water content from a known final water content (\S \ref{subsec:InitialWaterPredict}), efficiency of oxygen production through escape (\S \ref{subsec:O2Efficiency}), and a comparison to existing work with more comprehensive interior modeling \citep{krissansen2022predictions} to contextualize the atmospheric escape calculations of both studies (\S \ref{subsec:JoshKTCompare}).

\subsection{Escape Through Time} \label{subsec:EscapeTime}

\begin{figure*}
    \centering
    \includegraphics[width=\textwidth]{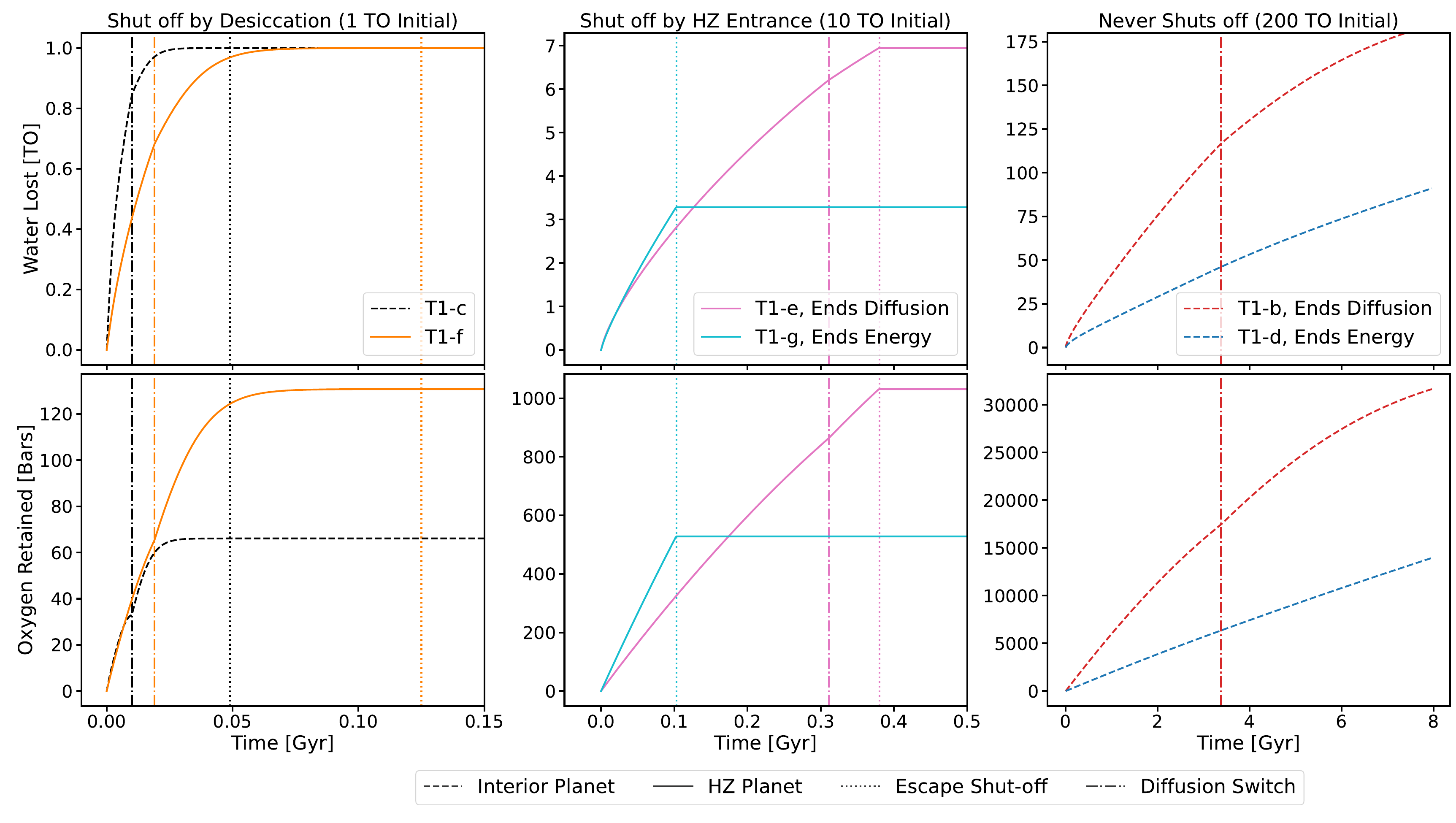}
    \caption{Illustration of the possible pathways the TRAPPIST-1 planets can take through escape physics in our model. Interior planets are denoted by dashed curves and habitable zone (HZ) planets are denoted with solid curves. Escape shut off times are shown by vertical dotted lines and times of switches from energy-limited to diffusion-limited escape are shown by vertical dot-dashed lines. \textbf{Left Column:} Escape is halted by complete planetary desiccation, for an interior planet, T1-c (black), and for a HZ planet, T1-f (orange), both with 1 TO initial water inventory. In these cases the atmospheric water is removed entirely, and the planet is last in diffusion-limited escape before the simulation halts. \textbf{Middle Column:} Simulations of HZ or outer planets where escape is halted by entrance to the HZ and the subsequent formation of a cold trap, all for initial water inventories of 10 TO. In this case escape could end in either the diffusion-limited (for T1-e, pink) or energy-limited (for T1-g, teal) regime. \textbf{Right Column:} Escape never shuts off as the planetary water reservoir is sufficiently large to sustain ongoing escape through the systems lifetime. This end state can only occur for interior planets (and results for T-1 b and d with an initial water inventory of 200 TO are shown here) as planets in or beyond the HZ have escape shut off by the formation of a cold trap. Once again, escape could end in either the diffusion-limited (for T1-b, red) or energy-limited (for T1-d, blue) regime, depending on total received insolation and initial water inventory.  \textbf{Top Row:} Shows water lost (in TO) over time for each of the column cases previously described. \textbf{Bottom Row:} Shows oxygen produced (in bars) over time for each of the column cases previously described.}
    \label{fig:EscapePathways}
\end{figure*}

Here, we describe the 6 distinct evolutionary pathways that the TRAPPIST-1 planets displayed in our model, due to combinations of escape physics, assumed initial water, and the interactions between the host star evolution and planetary properties (Figure \ref{fig:EscapePathways}). A given simulation's behavior as the host star luminosity and XUV evolves is primarily contingent on initial water content and orbital distance, but noticeable effects can also be attributed to the planet's mass and radius. Understanding the possible mechanisms of escape a planet could experience as a function of stellar evolution and planetary properties helps predict likely observables. \PeerReview{Note, that "Oxygen Retained"  (used on the y-axis label of Figures \ref{fig:EscapePathways}, \ref{fig:StatPlotInterior} and \ref{fig:StatPlotHZ}) refers to the oxygen that was produced and retained through the photolysis and thermal escape of water, when considering loss through hydrodynamic drag as the only oxygen sink.}

There are 3 possible states a planet may be in when the hydrodynamic escape phase ends, here termed ``end states": hydrodynamic thermal escape is halted by complete desiccation (Figure \ref{fig:EscapePathways}, left column), escape is halted by sufficient decrease in incident XUV marked by entrance to the habitable zone (HZ) (Figure \ref{fig:EscapePathways}, middle column), or escape never ends and is active at present day (Figure \ref{fig:EscapePathways}, right column). Within these cases, we can also consider whether the planet experienced diffusion-limited escape, or if it remained in the energy-limited regime for the duration of the simulation. As diffusion is responsible for mass fractionation, which may be observable \citep{Lincowski2019HDO}, knowledge of whether or not a planet may have experienced diffusion-limited escape is important for predicting the likelihood of fractionation signals in the planetary spectrum.

Planetary desiccation is due to an interplay between the planet's initial water content and its accumulated stellar insolation.  Desiccation may therefore be experienced by any planet in the system regardless of orbital distance, as those planets with lower initial water inventories are more likely to become desiccated, in addition to those with larger stellar insolation. If a planet becomes desiccated, it will pass the point where the mixing ratio of water is less than that of free oxygen (i.e., the ``diffusion switch''); thus, by definition planets that become desiccated will last be in the diffusion-limited escape regime.

As an M dwarf's luminosity fades with time, a planet once interior to the habitable zone may enter it \citep{LugerBarnes2015extreme}, and it is assumed that this will shut off hydrodynamic escape due to the formation of a cold trap that prevents sufficient water from reaching the exobase \citep{kasting1988runaway}. Only the outer planets (beyond TRAPPIST-1 d) will enter the HZ and experience this shut-off mechanism, thus it is primarily dependent on orbital position. We find the inner edge of the HZ reaches planets T1-e, f, g, and h at roughly $\sim$380 Myr, $\sim$170 Myr, $\sim$100 Myr, and $\sim$50 Myr, respectively, thus there is a short amount of time for the planets to lose water relative to the age of the system. Consequently, depending on their initial water inventories, these planets may or may not have experienced diffusion-limited escape, with lower initial water contents more likely to reach the diffusion switch before HZ entrance.

In the last end state, escape is never halted before the age of the system is reached, and the planet retains water and active escape up to present-day. Since the planets beyond TRAPPIST-1 d will have entered or passed beyond the habitable zone by the present day, simulations that end with active escape can only occur for interior planets. This end state is primarily dependent on initial water and typically only occurs for large initial water contents of $>$30 -- 60 TO, depending on orbital distance. Planets that follow this trajectory can experience diffusion-limited escape or stay in the energy-limited regime for the entire simulation, which is also dependent on initial water content and integrated insolation over time.

%To summarize, interior planets (T1-b -- d) can either end their simulation evolution by becoming desiccated, or by reaching the age of the system with water still present and escape active; the initial water content dividing these two end states is approximately 30 -- 60 TO depending on orbital distance. Outer planets (T1-e -- h) can either end their simulations by becoming desiccated in the case of low initial water contents $\leq$1 -- 2 TO, or by entering the habitable zone and forming a cold trap. The outermost planet, T1-h, is a special case that is not desiccated even for our lowest initial water content (1 TO), it additionally never reaches diffusion-limited escape as it passes through the HZ so quickly.
As uncertainty in the stellar evolution can affect what pathway a particular planet takes, and uncertainty in planetary mass and radius can sway borderline cases, this work leverages over 700,000 simulations to explore the full uncertainty space. The predicted final states of the planets and associated uncertainty from the culmination of all simulations is described in the following section. 

\subsection{Statistical Outcomes of Simulation Final States} \label{subsec:Statistical}

\begin{figure*}
    \centering
    \includegraphics[width=\textwidth]{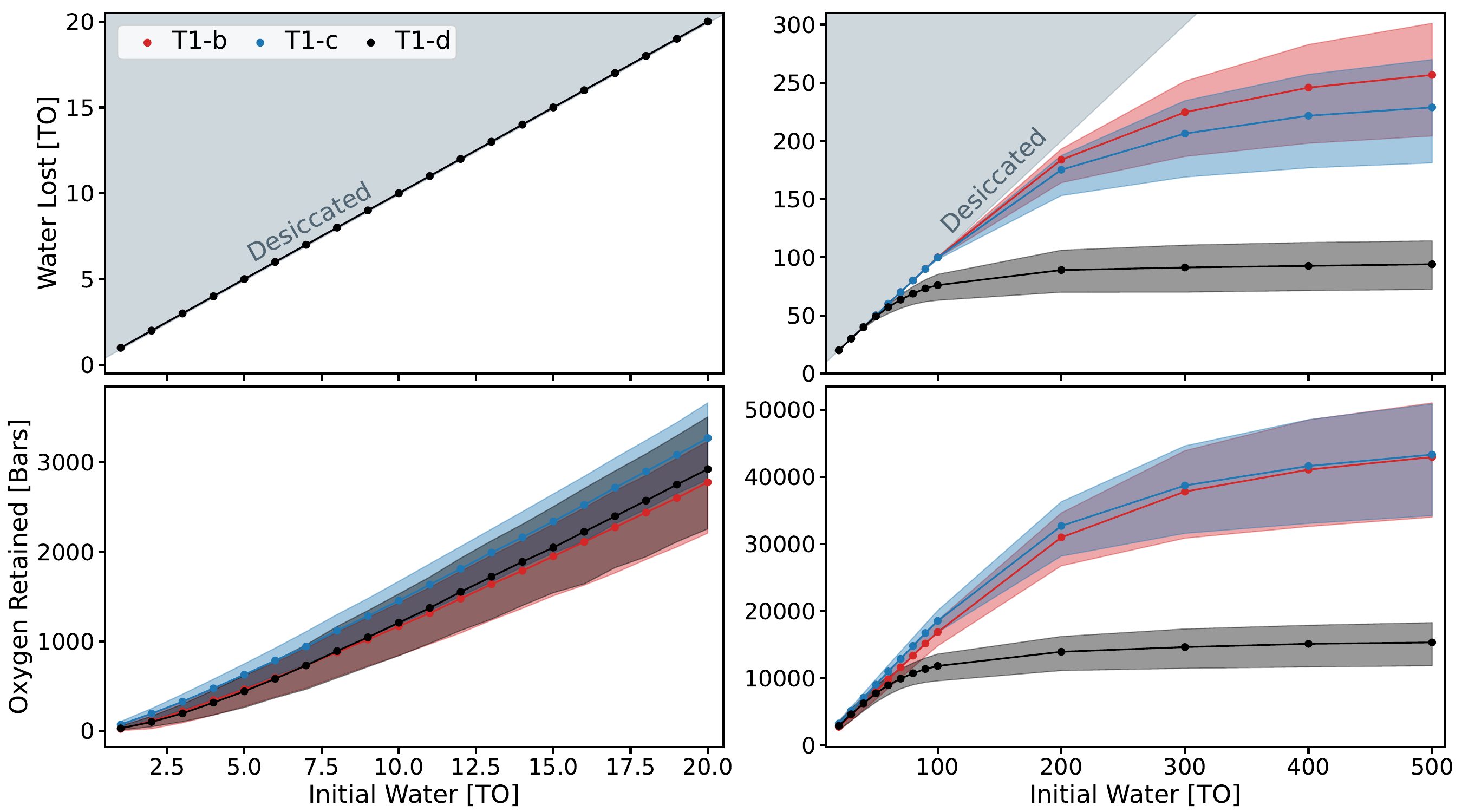}
    \caption{The predicted final water loss and final oxygen produced and retained as a function of initial water content for the TRAPPIST-1 interior planets T1-b (red), c (black), and d (blue). \textbf{Top Row:} Water lost (in TO) versus initial water (in TO); the shaded blue-grey region indicates where desiccation occurs. For water contents in the upper left panel, all three planets are desiccated and thus their curves are overlapping. \textbf{Bottom Row:} Oxygen produced and retained (in bars) versus initial water. \textbf{Left Column:} Low initial water contents (1 -- 20 TO). \textbf{Right Column:} High initial water contents (30 -- 500 TO). Shaded regions give the 1$\sigma$ uncertainty ranges (set by the 16$^{th}$ and 84$^{th}$ percentile values of the final state distributions). }
    \label{fig:StatPlotInterior}
\end{figure*}

\begin{figure*}
    \centering
    \includegraphics[width=\textwidth]{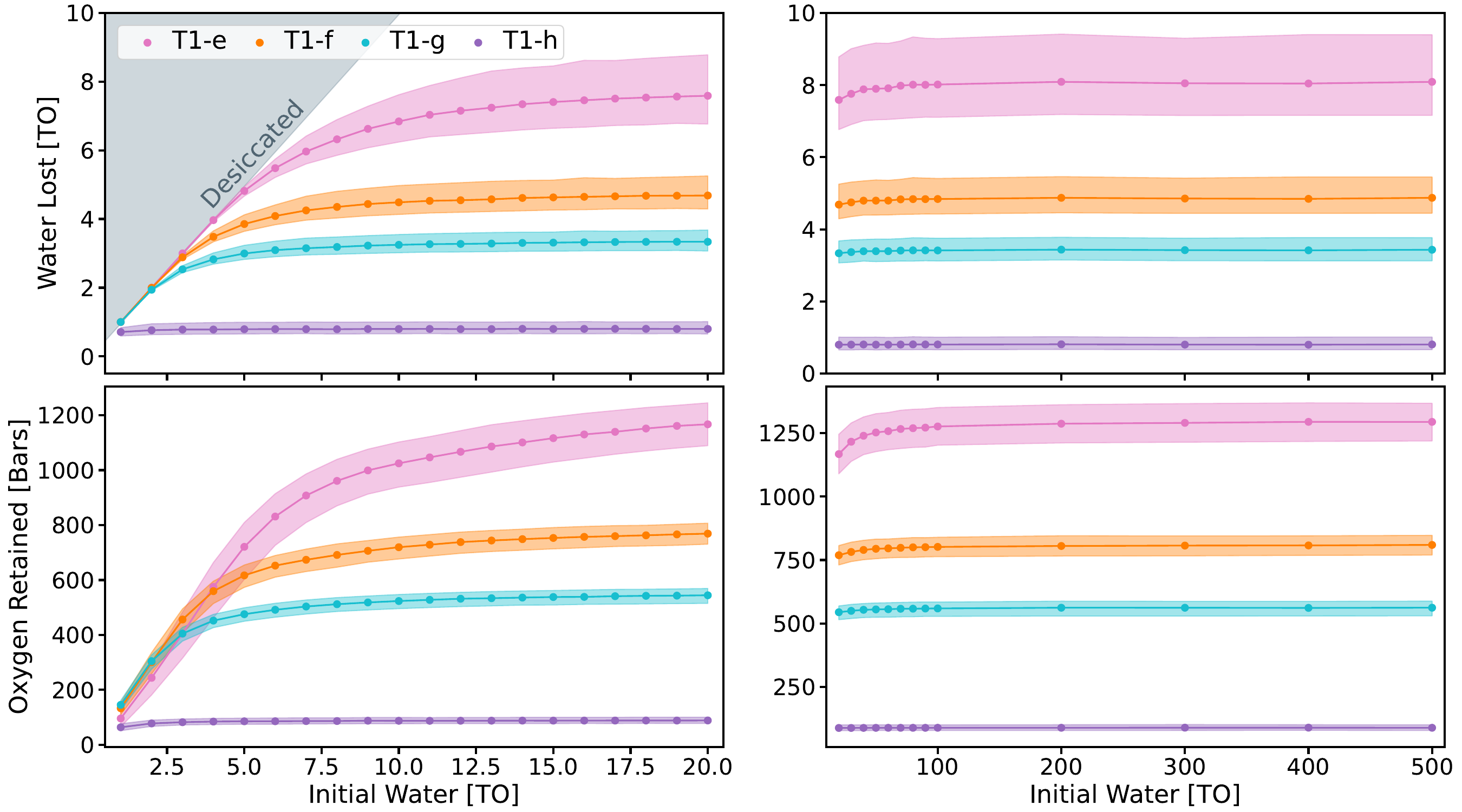}
    \caption{The predicted final water loss and final oxygen produced and retained as a function of initial water content for the TRAPPIST-1 habitable zone and outer planets T1-e (pink), f (orange), g (teal), and h (purple). The layout of the figure is the same as Figure \ref{fig:StatPlotInterior}.}
    \label{fig:StatPlotHZ}
\end{figure*}

Here we present the distributions from all simulations of water loss and oxygen production for each planet as a function of initial water content (Figures \ref{fig:StatPlotInterior} and \ref{fig:StatPlotHZ}). Figure \ref{fig:StatPlotInterior} shows the predicted final water loss (top row) and oxygen retained (bottom row) for the TRAPPIST-1 interior planets (T1-b, c, and d). For low initial water contents (1 -- 20 TO, left column), the interior planets are completely desiccated, enabling large amounts of oxygen production; T1-c retains the most oxygen against atmospheric escape, followed by d and then b. At larger water contents, T1-c continues to retain the most oxygen, but T1-b begins to retain more than T1-d due to T1-b's greater cumulative water loss at large initial reservoirs. T1-b, c, and d begin to avoid desiccation and retain fractional amounts of an ocean at initial water contents of 60, 50, and 30 TO, respectively. 

For higher initial water contents (30 -- 500 TO, right column), the interior planets' water loss curves begin to flatten, approaching a maximum possible water loss value. Maximum possible water loss here refers to the maximum amount of water a planet can lose given its maximum possible simulation time (the system age for interior planets, the time to the habitable zone for exterior planets) and sufficiently large water contents, such that adding more initial water would negligibly affect the predicted water loss. As water loss directly correlates with oxygen production, we also see maximum values for oxygen production and retention after thermal escape corresponding to maximum possible water loss. T1-d clearly displays this flattening off behavior indicating a maximum possible water loss of $\sim$90 TO and oxygen production of $\sim$15,000 bars. The continuing upward trend of water loss and oxygen production for T1-b and c suggest that the maximum possible water loss is first achieved for initial contents larger than those considered in this work ($>$500 TO), but we can conclude the maximum possible water loss of T1-b and c is greater than $\sim$250 TO and $\sim$230 TO, respectively. 

%For larger initial water contents, T1-c continues to retain the most oxygen. However, T1-b begins to retain more oxygen than T1-d due to T1-b's greater cumulative water loss at large initial reservoirs.
When considering only the balance of oxygen production and loss through thermal escape of water atmospheres, we may quantitatively explore the tendency of the planets to retain the oxygen they produce, or become completely airless. In some simulations with very low initial water, oxygen drag may be so efficient that only a negligible ($<$1 bar) atmosphere remains post-escape, even without the consideration of additional oxygen sinks \citep[e.g.,][]{schaefer2016predictions}; for an initial content of 1 TO, $\sim$19\%, $\sim$2.0\%, and $\sim$2.5\% of T1-b, c, and d simulations, respectively, were completely airless. This number drops to less than 1\% for all planets by an initial content of 5 TO.

Figure \ref{fig:StatPlotHZ} shows the water loss and oxygen production results for the HZ and outer planets (T1-e, f, g, and h) with the same layout as Figure \ref{fig:StatPlotInterior}. The outer planets rarely or never experience desiccation above a threshold initial water inventory; T1-e is desiccated for initial water up to 2 TO, while T1-f is only desiccated for 1 TO initial. T1-g is nearly desiccated for 1 TO, only retaining on the order of 10$^{-5}$ TO, but T1-h is never predicted to be desiccated for the initial water contents considered here ($\ge$1 TO).

% The outer planets have only a short time frame (approximately 50-380 Myr) before the star dims sufficeintly for them to enter the HZ, at which time thermal hydrodynamic escape shuts off, significantly limiting their maximum possible water loss, and oxygen buildup and retention. 
As HZ entrance halts escape on the outer planets, there is only a short time frame prior to entering the HZ (approximately 50-380 Myr) in which these planets can experience massive water loss through thermal hydrodynamic escape, which significantly limits their maximum possible water loss, and oxygen buildup and retention. T1-e, f, g, and h display maximum possible water losses of 8.0$^{+1.3}_{-0.9}$, 4.8$^{+0.6}_{-0.4}$, 3.4$^{+0.3}_{-0.3}$, and 0.8$^{+0.2}_{-0.1}$ TO, respectively, with corresponding maximum possible oxygen retention amounts of 1290$^{+75}_{-75}$, 800$^{+40}_{-40}$, 560$^{+30}_{-25}$, and 90$^{+10}_{-10}$ bars, respectively. For initial water inventories above 4 TO, water loss amounts tend to decrease with increasing orbital distances, with T1-e losing the most, followed by f, g, and h. However, for low initial water contents ($\le$2 TO), the planets lose comparable amounts of water, but T1-g retains the most oxygen due to its lower incident radiation followed by f, e, and then h. At an initial water inventory of 3 TO, T1-g moves to retain less oxygen than T1-f and e, and by 4 TO the ordering of the planets' oxygen retention tends back towards their orbital distance ordering as the closer planets lose more water overall and thus build-up more oxygen. 

Unlike the interior planets, the outer planets are rarely found to lose all produced oxygen. At an initial content of 1 TO, less than 1\% of simulations for all outer planets retain $<$1 bar of oxygen. For all 7 planets, the differences in oxygen retention ordering are influenced by the XUV environment and planetary gravity. In Section \ref{sec:Discussion}, we discuss these water loss (\S \ref{subsec:WaterlossDiscuss}) and oxygen build-up (\S \ref{subsec:OxygenDiscuss}) results, along with the broader implications and importance.  

\subsection{Predicting Initial Water Content} 

A fundamental limitation to our understanding of planetary evolutionary processes is that the atmospheric histories of exoplanets are not directly observable; however, they may be inferred from other measurable quantities. \PeerReview{Here, we show two different approaches to inferring initial water content: first by using predictions of the present-day water mass fraction of the planets \citep{agol2021refining} (\S \ref{subsec:InitialWaterPredict}), and then by using constraints on atmospheric oxygen content of the interior two planets inferred from the most recent JWST observations \citep{Greene2023T1bJWST,Zieba2023T1cJWST} (\S \ref{subsubsec:JWSTpredictresults}).}

\subsubsection{\PeerReview{Using Present-day Water Mass Fractions}}\label{subsec:InitialWaterPredict}

Within our model and its limitations, we have shown a direct relationship between the initial and final water contents of the TRAPPIST-1 planets (Figures \ref{fig:StatPlotInterior} \& \ref{fig:StatPlotHZ}); thus, it is possible to take a known or predicted final water content and determine the initial water content our model predicts that is most likely to produce that final state. As our models only consider one evolutionary process (atmospheric escape driven by evolving stellar conditions), we caution that this analysis is meant to be a proof of concept. Our results are dependent on our model assumptions and limitations, as well as the assumptions made in the final state predictions we adopt.

%To estimate the final water content for each of the TRAPPIST-1 planets, 
In this analysis, we use the present-day water mass fraction predictions for each of the TRAPPIST-1 planets from \citet{agol2021refining}, which are derived from observationally constrained planetary densities, assuming an Earth-like core mass fraction of 32.5\%. To calculate these water mass fractions, \citet{agol2021refining} assumes all water on the planets is on the surface; steam \citep{Turbet2020steamwatermodel} and condensed \citep{Dorn2017condensedwatermodel} water mass-radius relationships were used for the interior (T1-b -- d) and outer (T1-e -- h) planets, respectively. As any surface water on the interior planets is expected to be vaporized \citep{Turbet2019runawayradiusinflation,Turbet2020steamwatermodel}, they conclude the interior planets should have water mass fractions drastically less than 0.01 wt\%. Since a broad range of our initial water contents would result in this desiccated state for the inner planets, we cannot constrain their most probable initial water contents from this present day desiccated state alone; thus, we only attempt to constrain initial water abundance for the outer planets. The predicted present-day water constraints we adopt for the outer planets T1-e, f, g, and h are given in Table \ref{tab:wmfpresentday} \citep[from Table 9 in ][]{agol2021refining}. 

\begin{table}[h]
    \centering
    \hspace*{-1cm}\begin{tabular}{c|c}
        Body & Water Mass Fraction [H$_{2}$O wt\%] \\
        \toprule
        T1-e & 2.9$^{+1.7}_{-1.5}$ \\
        \midrule
        T1-f & 4.5$^{+1.8}_{-1.2}$ \\
        \midrule
        T1-g & 6.4$^{+2.0}_{-1.6}$ \\
        \midrule
        T1-h & 5.5$^{+4.5}_{-3.1}$ \\
    \end{tabular}
    \vspace{2mm}
    \caption{Present-day water mass fraction constraints for the outer planets (T1-e, f, g, and h) that we use to predict most probable initial water content in our model. \PeerReview{These constraints are taken from \citet{agol2021refining} when assuming an Earth-like CMF of 32.5\%, and are not definitively the true water mass fractions of the planets.}}
    \label{tab:wmfpresentday}
\end{table} 

\begin{figure}
    \centering
    \includegraphics[width=\textwidth]{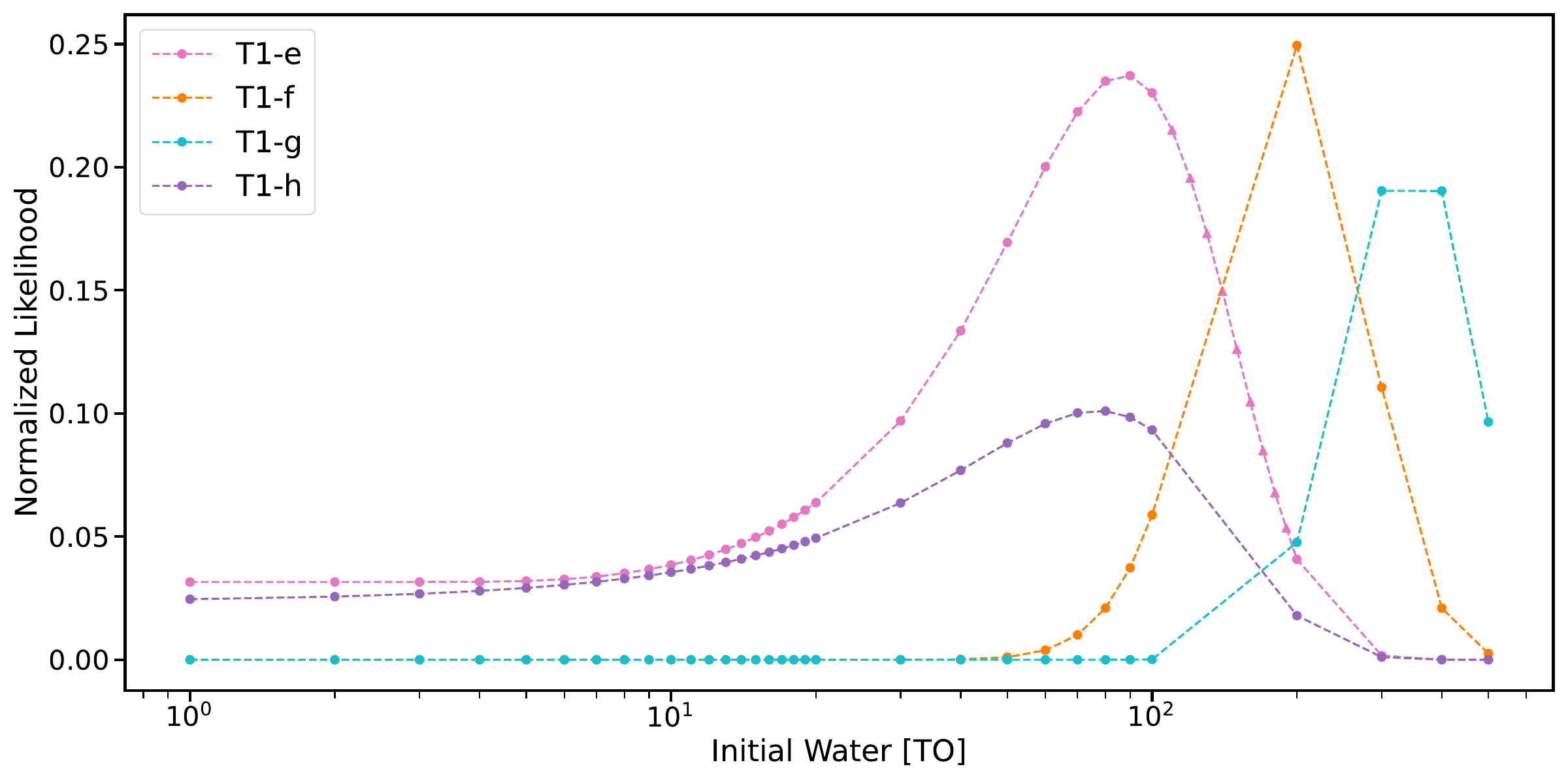}
    \caption{The likelihood of each initial water content (in TO) needed to reproduce the predicted present-day water contents from \citet{agol2021refining} for each of the outer planets. The peaks denote the initial water content on our sampling grid most likely to produce the final content in our predictions. An additional 9 simulation suites were ran for T1-e only, between 100 -- 200 TO to fill out a smooth, continuous curve\PeerReview{, and results from these additional simulations  are denoted by triangle markers.}}
    \label{fig:PDFSumsHZPlanets}
\end{figure}

The method described in Appendix \ref{appendix:initialpredictionMethods} is used to determine the initial water content most likely to produce these predicted final water contents in our model. Briefly, the present-day water predictions are first modeled with a gamma probability density function (PDF); we then sum the individual simulations' values on that PDF and normalize for the number of simulations per initial water content to get a likelihood of that initial content given our predicted final state. Figure \ref{fig:PDFSumsHZPlanets} shows the likelihood versus initial water content per planet. The peaks of these curves indicate the initial water content on our sampling grid that best recreated the predicted final states; for T1-e, f, g, and h these are 90, 200, 300, and 80 TO, respectively.

\begin{figure}
    \centering
    \includegraphics[width=0.7\textwidth]{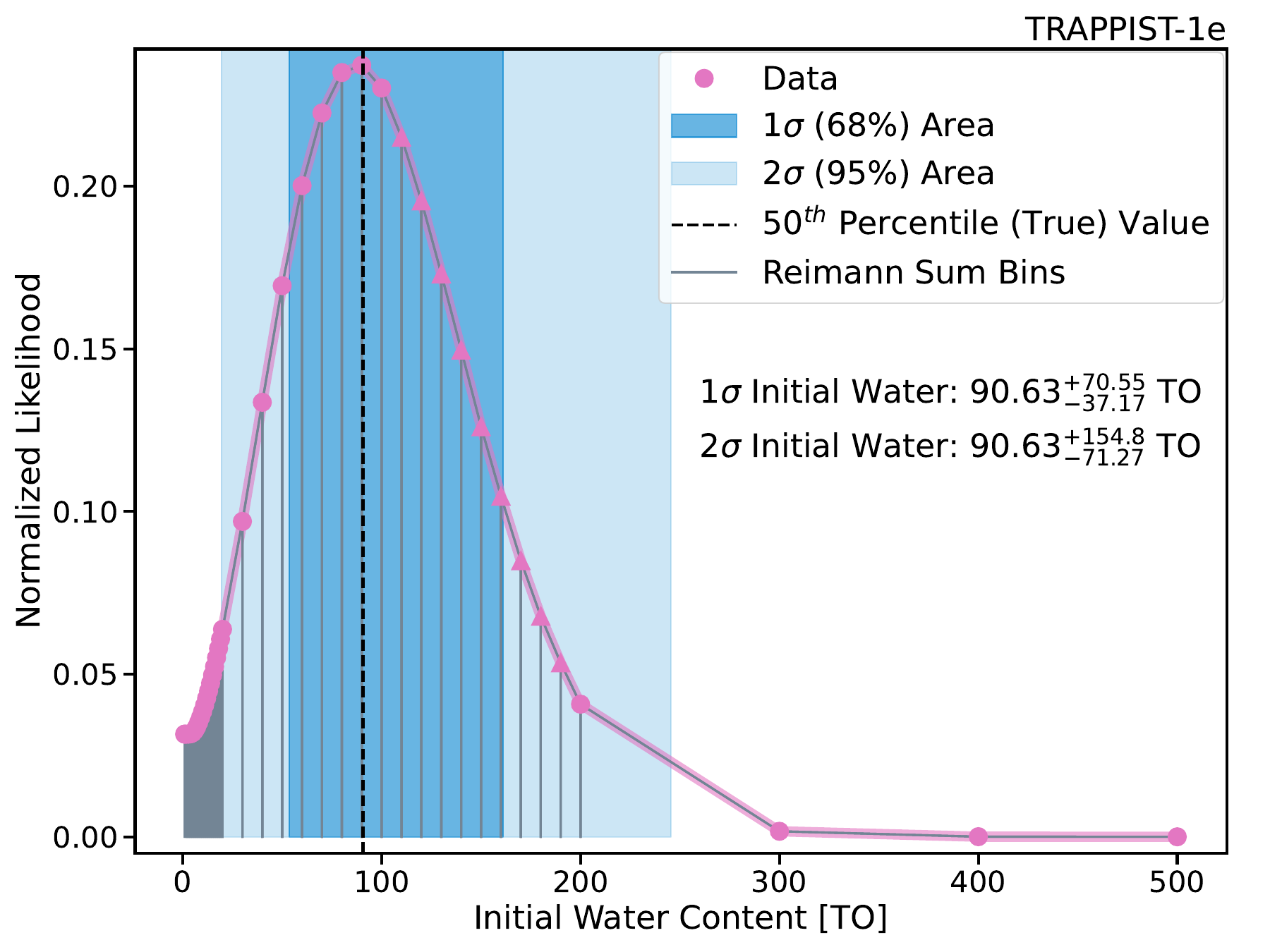}
    \caption{An illustration of the method used to determine the most probable initial water content for T1-e that would reproduce our predicted present-day content from \citet{agol2021refining} (Table \ref{tab:wmfpresentday}). Same layout as Figure \ref{fig:PDFSumsHZPlanets} but with linear x-axis scaling; the y-axis is the likelihood that  each initial water content (in TO; x-axis) produces the desired final state. The data points that were simulated are shown by the pink points, the bins used in the Riemann sum are shown in grey, the determined 50$^{th}$ percentile or true predicted value is shown by the vertical dashed black line, and the 1$\sigma$ and 2$\sigma$ confidence intervals are shown by the shaded dark and light blue regions, respectively. The low initial water contents (1 -- 20 TO) are much more finely spaced (step size of 1 TO) which is why the Riemann sum bin widths are not resolvable there. Lines connecting data points are used for the tops of the Riemann sum bins.}
    \label{fig:T1eRiemann}
\end{figure}

We further derive the 50$^{th}$ percentile value of probable initial water with uncertainty for T1-e, as a proof of concept. To find the most probable initial water content predicted by our model with associated confidence intervals, we need to interpolate across the curves in Figure \ref{fig:PDFSumsHZPlanets}, which must possess enough data be smooth and continuous. Many of the planets we've considered do not possess smooth and continuous curves, and adding more data to these requires many additional simulation suites and computational expense; because of this, and the inherent uncertainty in our predictions, we complete this additional analysis for T1-e only as proof of concept. To obtain a continuous curve for T1-e, an additional 9 simulation suites were completed for initial water contents between 100 -- 200 TO (step size of 10 TO)\PeerReview{; additional data points for T1-e are shown by a triangle marker in Figures \ref{fig:PDFSumsHZPlanets} \& \ref{fig:T1eRiemann}}. These additional simulation suites are only considered in this initial water content analysis, as their inclusion in the results presented in Section \ref{subsec:Statistical} would not change the conclusions.

The interpolation for T1-e is accomplished with a modified Riemann sum, illustrated by Figure \ref{fig:T1eRiemann}, to determine the 50$^{th}$ percentile value of predicted initial water content and associated 1$\sigma$ and 2$\sigma$ confidence intervals (see Appendix \ref{appendix:initialpredictionMethods} for mathematical description). The 1$\sigma$ confidence interval of most probable initial water content to produce our present-day water content prediction for T1-e is 90.6$^{+70.6}_{-37.2}$ TO, and the 2$\sigma$ confidence value is 90.6$^{+155}_{-71.3}$ TO. Applications and importance of this method are further discussed in Section \ref{subsec:PredictingInitialDiscuss}.

\subsubsection{\PeerReview{Using Constraints on Atmospheric Oxygen Abundance}} \label{subsubsec:JWSTpredictresults}

\PeerReview{Although the statistical inference technique (\S\ref{subsec:InitialWaterPredict}) can produce an upper limit on initial water abundance, a more refined estimate can be obtained by deriving tighter constraints from existing and future observations of atmospheric oxygen abundance on T1-b and c. Recently, \citet{Greene2023T1bJWST} used JWST/MIRI to obtain a blackbody brightness temperature of $T_{B}$ = 503$^{+26}_{-27}$ K for T1-b from secondary eclipse observations taken at 15 $\mu$m, which they find to be an excellent match to the 508K theoretical prediction of dayside temperature for an airless world with no heat redistribution.  If a significant atmosphere was present, then the subsequent redistribution of heat from the dayside  to nightside, and/or absorption from atmospheric CO$_2$, would result in an average dayside temperature at 15 $\mu$m that is significantly lower than that observed. \citet{Greene2023T1bJWST} therefore conclude that T1-b likely does not have a substantial atmosphere today. This observation is also consistent with nearly half of all evolutionary model runs from \citet{krissansen2022predictions}, which predict complete atmospheric erosion on T1-b across their considered range of initial water (0.7 -- 300 TO).}

\PeerReview{The observational constraint that T1-b is likely airless today can be used with our grid of atmospheric evolutionary outcomes to identify limits on the planet's initial surface water. Considering only the apparent lack of water, our model predicts the planet must have begun with $\le$60 TO to be in a desiccated present-day state. Although water does not remain in these simulations, an oxygen atmosphere could still be supported, especially for the higher end of the initial water range, and depending on assumed sinks. However, a  dense O$_2$ atmosphere, with even a small amount of outgassed CO$_2$, could produce an absorption feature at 15$\mu$m that is also precluded by the observational constraints \citep{Greene2023T1bJWST,Ih2023t1binterpret,Lincowski2023T1c}. The tighter observational constraint that T1-b is currently desiccated and \textit{also} likely does not have an oxygen atmosphere, combined with considerations of oxygen sinks over the lifetime of the system, will significantly lower this upper limit on initial water inventory.}

\PeerReview{Though detailed oxygen surface sink modeling is outside the scope of the current study, to give an example of how one may further constrain this upper limit we estimate a few oxygen sinks here, first considering only oxygen loss from the magma ocean and hydrodynamic escape while water loss is ongoing. Assuming magma ocean removal operates on the same timescales as hydrodynamic escape \citep{Hamano2013typesmagmaoc}, we estimate a removal of 3 TO of oxygen to the magma ocean ($\sim$630 bars on T1-b), based on studies of the early Venusian environment and theoretical studies on magma ocean evolution for the TRAPPIST-1 planets \citep{Gillmann2009magmaremoval,schaefer2016predictions,Wordsworth2018redox,barth2021magma}; and from our simulations we find 4 TO equivalents of oxygen would be lost to hydrodynamic drag. These combined give maximum initial surface water contents of 7 TO, to be consistent with an airless present day T1-b.}
\PeerReview{However, loss of atmospheric oxygen can still continue after total loss of the water inventory, via processes such as non-thermal escape and dry crustal oxidation.  Following \citet{krissansen2022predictions}, we consider a non-thermal escape parameterization of $\sim$100 bars of total oxygen loss across the system's lifetime, 
%\citep[as was done in ][]{krissansen2022predictions}
and an average estimate of oxygen loss through dry crustal oxidation of $\sim$10 Tmol/year ($\sim$55 bars/Gyr) after magma ocean solidification \citep[following the perpetual runaway greenhouse scenario in][]{Krissansen2021oxygenfalsepos}. Given a system age of $\sim$8 Gyr \citep{birky2021improved} and initial water contents of around 15 TO, T1-b may endure hydrodynamic escape for up to 1 Gyr, leaving 7 Gyr for dry crustal oxidation; thus, we conservatively estimate 385 bars of total oxygen removal through dry crustal oxidation. With these two additional oxygen sinks (magma ocean removal and dry crustal oxidation), our upper limit constraint on initial surface water, consistent with complete atmospheric erosion on T1-b at the present day, would be lowered to 12 TO. However, note that with more rigorous modeling of oxygen removal, this upper limit would likely be higher. Overall we can conclude the upper limit on initial water content as constrained by observations of T1-b \citep{Greene2023T1bJWST} is between 12 -- 60 TO.}

\PeerReview{Although the above upper limit constraints on initial water abundance are developed with the observational constraint from TRAPPIST-1b, habitability studies of the system may be more concerned with the lower limit of initial water, which could inform the minimum amount of water the HZ planets possess at present-day. Constraining this lower limit may be possible by considering what we know of the atmospheric state of TRAPPIST-1c in addition to that of T1-b. Recent T1-c secondary eclipse observations with JWST/MIRI show a brightness temperature of $T_{B}$ = 380$\pm$31 K \citep{Zieba2023T1cJWST}. This is 50K below, and 1.7-$\sigma$ away from, the temperature expected for an airless body with zero heat redistribution (430 K). It is also 1.6-$\sigma$ from an ultramafic surface with the strong weathering expected for this intensely UV-irradiated world \citep{Zieba2023T1cJWST}.  The T1-c JWST 15$\mu$m data are best fit by either an airless world with a high albedo surface, or a thin $<$0.1 bar O$_{2}$-dominated atmosphere \citep{Zieba2023T1cJWST,Lincowski2023T1c}. However, other atmosphere types are still consistent with the data, including a 3 bar steam atmosphere, which is within 1.7-$\sigma$ of the observed secondary eclipse value \citep{Lincowski2023T1c}.}

\PeerReview{Given their different stellar distances, similar planetary properties and inferred geologic oxygen sinks, there is a narrow range of initial surface water contents that will simultaneously produce an airless T1-b and non-negligible oxygen on T1-c, even if both planets are found to be desiccated.  As T1-b and T1-c are similarly interior to the HZ, possess comparable surface gravities ($g_{T1c} \sim 0.99 g_{T1b}$), and likely formed with a similar composition ($\rho_{T1c} \sim 1.004 \rho_{T1b}$) they likely experience comparable geologic and escape evolution. However, we have shown that since T1-c is slightly farther away from the host ($a_{T1c} \sim 1.37 a_{T1b}$), it will retain more oxygen for equivalent amounts of water loss as T1-b due to the lower incident XUV flux (see \S \ref{subsec:O2Efficiency} and Figure \ref{fig:RetentionEfficiency}).}

\PeerReview{Considering oxygen removal through a magma ocean, dry crustal oxidation, and non-thermal escape, we run an Markov chain Monte Carlo (MCMC) to constrain the initial water content that would produce both an airless T1-b and a tenuous 0.1 bar O$_{2}$-dominated atmosphere on T1-c, consistent with the currently available JWST data \citep{Greene2023T1bJWST,Zieba2023T1cJWST,Ih2023t1binterpret,Lincowski2023T1c}. To conduct the MCMC, we use the publicly available \textbf{emcee} model \citep{ForemanMackey2013emcee}. MCMC walkers fit for the initial water content by running VPLanet \citep{Barnes2020VPLanet} simulations of T1-b and T1-c that predict the oxygen build-up of the system. After a simulation, the estimated oxygen removal is subtracted, and the MCMC checks to see that T1-b has no oxygen remaining while T1-c is compared to a 0.1 bar oxygen atmosphere with an uncertainty of 10 bars, as a 10 bar oxygen atmosphere begins to show $>$3$\sigma$ disagreement from the 15 $\mu$m measurement of T1-c \citep{Zieba2023T1cJWST}. As described earlier in the section, we estimate removal of 3 TO-worth of oxygen from a magma ocean \citep{Gillmann2009magmaremoval,barth2021magma}, 100 bars from non-thermal escape \citep{krissansen2022predictions}, and 10 Tmol/Gyr ($\sim$55 bars/Gyr) from dry crustal oxidation after the hydrodynamic escape period \citep{Krissansen2021oxygenfalsepos}. We vet convergence by considering the chain's autocorrelation time; in total, we run 20 walkers for 80,000 steps with 1,000 burn-in steps (for a total of $>$1.5 million VPLanet simulations), this results in a chain length that is several times the autocorrelation time, suggesting convergence has been achieved \citep{ForemanMackey2013emcee}. We note that our oxygen removal parameterizations are meant to be low estimates of the total oxygen removal the planets may experience, as we are searching for a minimum initial surface water mass and greater oxygen removal would serve to raise the minimum.}

\begin{figure}
    \centering
    \includegraphics[width=\textwidth]{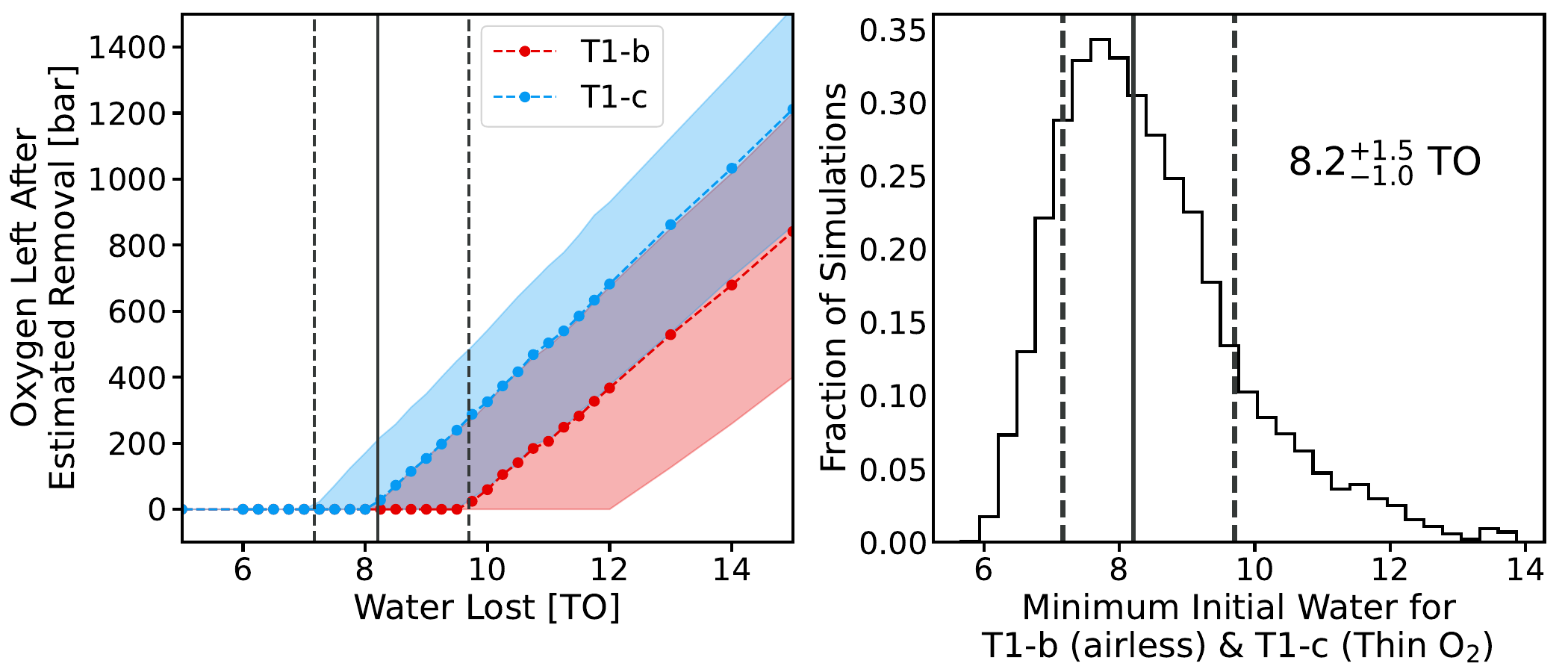}
    \caption{\PeerReview{\textbf{Left:} The oxygen left in the atmospheres of T1-b (red) and T1-c (blue) after an estimated removal of 3 TO worth of oxgyen from a magma ocean, 100 bars from non-thermal escape, and 10 Tmol/yr ($\sim$55 bars/Gyr) from dry crustal oxidation after the planet becomes desiccated (see text for estimation explanation). \textbf{Right:} The resulting histogram of initial water contents predicted by MCMC when fitting for the scenario that T1-b has lost all oxygen while T1-c retains 0.1 bar of atmospheric oxygen following the estimated removal. The vertical solid black lines on both panels denotes the 50$^{th}$ percentile value predicted by the MCMC while the vertical dashed black lines denote the 68\% (1$\sigma$) confidence interval. This analysis predicts a minimum initial water content of 8.2$^{+1.5}_{-1.0}$ TO to produce the desired final state, assuming both planets begin with the same initial water content. The amount of oxygen removal is meant to be an underestimate to produce a conservative lower limit, as additional oxygen removal would increase the minimum initial water required.}}
    \label{fig:JWSTAnalysis}
\end{figure}

\PeerReview{Figure \ref{fig:JWSTAnalysis} shows the results of our MCMC analysis; the left panel is an illustration of the atmospheric oxygen left after removal by the estimated oxygen sinks, and the right panel is the distribution of initial water contents reported by the MCMC that best fit an airless T1-b and a T1-c with a 0.1 bar oxygen atmosphere. With 68\% confidence (1$\sigma$) we find the minimum initial water content to be 8.2$^{+1.5}_{-1.0}$ TO. This value with associated uncertainty is shown by vertical lines on both panels of figure \ref{fig:JWSTAnalysis}.}

\PeerReview{As one last note, in considering the outer planets, under the assumptions of a similar magma ocean removal rate during hydrodynamic escape, followed by a removal of $\sim$135 bars/Gyr analogous to geological O$_2$ sinks on the modern Earth (including weathering, volcanism, and subduction) \citep{Catling2014Oxygensourcesandsinks}, 
%anoxic atmospheres 
atmospheres lacking in O$_2$ are possible for all initial water contents considered.
We point out, however, that these are rough estimates, and further work would be needed in tracking oxygen sinks over time to produce more accurate results. }

%we find initial surface water content is constrained between 7 -- 12 TO when considering the 1$\sigma$ uncertainty region of our results. When considering only the 50$^{th}$ percentile values we obtain the tighter constraint of 9 -- 10 TO initial water abundance. This estimated range is in agreement with \citet{Zieba2023T1cJWST}, who suggest an initial water of 9.5$^{+7.5}_{-2.3}$ TO through an analysis of conservative CO$_{2}$ upper limit constraints, albeit with a similar atmospheric escape model \citep{LugerBarnes2015extreme,schaefer2016predictions}. Thus, by considering the two planets together we predict a lower and upper limit constraint, which greatly improves our theoretical predictions from only an upper limit constraint when considering T1-b alone.

\subsection{Oxygen Retention Efficiency} \label{subsec:O2Efficiency}

%By quantifying oxygen retention efficiency, we evaluate the balance of thermal hydrodynamic escape to act as both a significant oxygen source and sink.  
Quantification of the oxygen retention efficiency allows us to evaluate the net balance between oxygen produced via water photodissociation and subsequent hydrogen escape, and oxygen loss due to drag during thermal hydrodynamic escape.  
%We therefore define oxygen retention efficiency as the percentage of oxygen produced through water photolysis that was retained by the planet following loss through drag:
Oxygen retention efficiency is given by:

\begin{equation}
    \label{eq:efficiency}
    \large
    \eta_{Retention} = \ddfrac{M_{O_{2},produced} - M_{O_{2},lost}}{M_{O_{2},produced}} \times 100,
\end{equation}
where $M_{O_{2},produced}$ is the total mass of oxygen that was produced by the cumulative hydrogen loss and $M_{O_{2},lost}$ is the mass of oxygen that was lost through drag during escape. Note that this only takes into account production and losses via escape mechanisms, and oxygen may be further lost through sinks not considered here \citep[e.g.,][]{schaefer2016predictions,Schaefer2017redox}.
%so this should be interpreted as the efficiency of oxygen retention through the hydrodynamic escape process. 

\begin{figure}
    \centering
    \includegraphics[width=\textwidth]{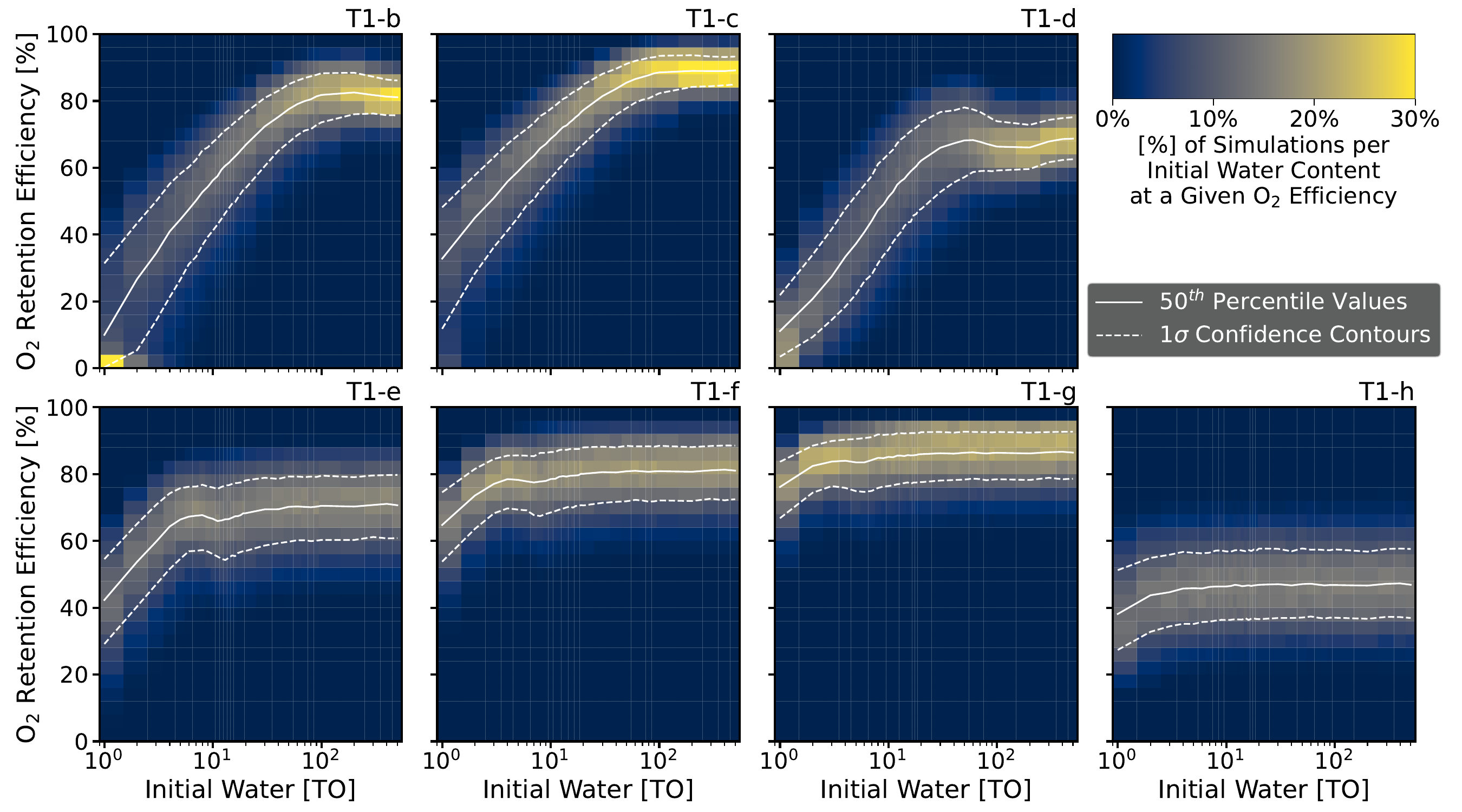}
    \caption{Oxygen retention efficiency as a function of initial water content in TO. \textbf{Top Row:} Interior planets, T1-b, c, and d. \textbf{Bottom Row:} Outer planets, T1-e, f, g, and h. Coloring shows the percentage of simulations at a particular initial water content that fell in a given oxygen retention efficiency bin (e.g., each column corresponds to one initial water content and sums to 100\%). Contours are given for the 50$^{th}$ percentile values (solid white lines) and the 1$\sigma$ confidence region (dashed white lines).}
    \label{fig:RetentionEfficiency}
\end{figure}

Figure \ref{fig:RetentionEfficiency} shows the oxygen retention efficiency (Eq. \ref{eq:efficiency}) as a function of initial water content with contours for the 50$^{th}$ percentile values and 1$\sigma$ confidence region for all 7 planets. The color scale of Figure \ref{fig:RetentionEfficiency} shows, for a particular initial water content, the percentage of simulations that had a given oxygen retention efficiency. Additionally, Table \ref{tab:OverallEfficiency} gives the 50$^{th}$ percentile values and 1$\sigma$ uncertainties for oxygen retention efficiency for our lowest (1 TO) and highest (500 TO) initial water case for all 7 planets. This result shows that oxygen retention efficiency increases with increasing orbital distance, initial water content, and planetary gravity. At low initial water contents, there is a strong distance dependence where efficiency increases with increasing orbital separation due to the drop in XUV flux; however, planets with low gravity may break this distance trend and show lower than expected \ce{O2} retention efficiency when compared to the distance trend, such as T1-d and h. High initial water inventories (100 TO or more) typically lead to larger retention efficiencies than those for lower water inventories, and these retention efficiencies display a weaker distance dependence; for example T1-c at its largest initial water content possesses the largest retention efficiency of any planet. This is because high initial water contents sustain escape to much later ages; as oxygen loss through drag will be higher early on with a higher XUV environment, decreasing with time as XUV decreases, the cumulative oxygen retention efficiency will be higher for planets that remain in active escape to later times. In comparison, planets where hydrodynamic escape shuts off earlier will have lower cumulative oxygen retention efficiency because all loss processes occur in a high XUV environment. 
%Higher water contents allow the interior planets to remain in the escape regime and generating oxygen to much later ages than the outer planets, which halt at HZ entrance; thus, the lower XUV environment at later system ages leads to higher oxygen retention efficiencies for the interior planets with large initial water contents.
%However, because the interior planets remain in the escape regime to much later times than the outer planets at high initial water contents, the efficiency results for the interior planets are influenced to be higher because incident XUV and loss through drag decrease with time.
This influence is further shown by the plateau of retention efficiency as planets reach initial water contents that lead to their maximum possible water loss (i.e., the maximum amount of water they may lose given the maximum possible simulation time and an effectively infinite initial reservoir), indicating that retention efficiency is directly dependent on time spent in the escape regime with evolving XUV conditions, which is in turn dependent on initial water. \PeerReview{At low initial water contents, the interior planets (T1-b, c, and d) retain $\sim$10\% -- 30\% of oxygen produced, while the outer planets retain $\sim$40\% -- 70\%. At high initial water contents, this number increases to $\sim$70\% -- 90\% for all planets in the system, excluding T1-h, which has extremely low gravity and loses $\lesssim$1 TO for any initial content, thus displaying retention efficiency that remains at $\sim$46\% even as initial water increases.}

\PeerReview{However, counter to the general trend of retention efficiency increasing with initial water content described above, we also find that there is a limited range of initial water contents for each planet where feedback from diffusion processes causes retention efficiency to \textit{decrease} with increasing initial water; this decreasing behavior is particularly noticeable for T1-d near $\sim$100 TO and T1-e near $\sim$10 TO in Figure \ref{fig:RetentionEfficiency}. For a particular planet, this decreasing behavior occurs when the initial water content approaches an amount large enough to prevent the planet from entering the diffusion-limited escape regime over its escape lifetime (either prior to entering the HZ, or up to present day). Since diffusion-limited escape prohibits oxygen loss and favors oxygen retention, avoiding a diffusion-limited escape period will decrease oxygen retention efficiency. Thus, when this regime is prevented entirely, we see a range where retention efficiency decreases with increasing initial water, as the influence of the lack of diffusion-limited escape outweighs the gain in efficiency from increasing the mixing ratio of water. However, this behavior only occurs for a narrow range of initial water abundances, and increasing efficiency with increasing initial water contents is seen on either side of that range.}

When losing equal amounts of water a strong distance dependence persists of increasing retention efficiency with increasing orbital distance, even when increasing initial water content; this behavior is apparent by the results in Table \ref{tab:1TOLossResults}, which show the efficiency of oxygen retention following an equal loss of 1 TO for a low (1 TO) and high (9, 10, or 11 TO) initial water case. Note, as this result compares an equal water loss, simulations that do not lose a full TO are not included. Consequently many cases for T1-g, and all cases for T1-h are excluded, as the former does not typically lose a full TO when it has a low starting inventory, and the latter does not typically lose a full TO for any starting inventory considered. Overall, decreases in orbital distance, initial water, and gravity will all decrease oxygen retention efficiency.

\begin{table}[h]
    \centering
    \hspace*{-1cm}\begin{tabular}{c|c|c}
         \multicolumn{3}{c}{Oxygen Retention Efficiency [\%] for Cumulative Water Loss} \\
        \midrule
        Body & Lowest Initial, 1 TO & Highest Initial, 500 TO \\
        \toprule
        T1-b & 9.979$^{+21.39}_{-9.725}$ & 81.11$^{+4.999}_{-5.410}$ \\
        \midrule
        T1-c & 32.79$^{+15.35}_{-21.01}$ & 89.14$^{+4.122}_{-4.221}$ \\
        \midrule
        T1-d & 11.11$^{+10.74}_{-7.668}$ & 68.71$^{+6.393}_{-6.157}$ \\
        \midrule
        T1-e & 42.31$^{+12.24}_{-13.17}$ & 70.68$^{+9.078}_{-9.853}$ \\
        \midrule
        T1-f & 64.78$^{+9.762}_{-10.98}$ & 81.04$^{+7.487}_{-8.562}$ \\
        \midrule
        T1-g & 76.14$^{+7.546}_{-9.347}$ & 86.43$^{+6.227}_{-7.829}$ \\
        \midrule
        T1-h & 38.13$^{+13.12}_{-10.76}$ & 46.87$^{+10.72}_{-9.938}$ \\
    \end{tabular}
    \vspace{2mm}
    \caption{The efficiency of oxygen retention [\%] after the cumulative water loss predicted by our model at the lowest (1 TO, middle column) and the highest (500 TO, right column) initial water content. Uncertainties reported are 1$\sigma$.}
    \label{tab:OverallEfficiency}
\end{table}

\begin{table}[h]
    \centering
    \hspace*{-1cm}\begin{tabular}{c|c|c}
         \multicolumn{3}{c}{Oxygen Retention Efficiency [\%] for a 1 TO Loss} \\
        \midrule
        Body & Initial 1 TO & Initial 9, 10, or 11 TO \\
        \toprule
        T1-b & 9.979$^{+21.39}_{-9.725}$ & 16.83$^{+14.31}_{-8.395}$ \\
        \midrule
        T1-c & 32.79$^{+15.35}_{-21.01}$ & 32.85$^{+11.90}_{-15.40}$ \\
        \midrule
        T1-d & 11.11$^{+10.74}_{-7.668}$ & 17.49$^{+6.754}_{-6.181}$ \\
        \midrule
        T1-e & 42.31$^{+12.24}_{-13.17}$ & 39.49$^{+10.33}_{-9.698}$ \\
        \midrule
        T1-f & 64.78$^{+9.762}_{-10.98}$ & 60.31$^{+10.46}_{-10.87}$ \\
        \midrule
        T1-g & -- & 73.12$^{+9.158}_{-10.69}$ \\
        \midrule
        T1-h & -- & -- \\
    \end{tabular}
    \vspace{2mm}
    \caption{The efficiency of oxygen retention [\%] following a 1 TO water loss. Uncertainties reported are 1$\sigma$. The middle column shows the results for planets with an initial inventory of 1 TO, and the right column shows the results for planets with an initial inventory of 9, 10, or 11 TO.}
    \label{tab:1TOLossResults}
\end{table} 

%%%%%%%%%%%%%%%%%%%%%%%%%%%% Mann-Whitney U Test values from previous result
%\begin{table}[h]
%    \centering
%    \hspace*{-1cm}\begin{tabular}{c|c|c}
%         \multicolumn{3}{c}{Significant P-values from the Mann-Whitney U Test} \\
%        \midrule
%        Body & Worse O$_{2}$ Retention at \newline 1 TO Initial & Better O$_{2}$ Retention at\newline 1 TO Initial \\
%        \toprule
%        T1-b & $3.11 \times 10^{-181}$ & -- \\
%        \midrule
%        T1-c & -- & -- \\
%        \midrule
%        T1-d & $0.0$ & -- \\
%        \midrule
%        T1-e & -- & $6.33 \times 10^{-34}$ \\
%        \midrule
%        T1-f & -- & $1.41 \times 10^{-118}$ \\
%    \end{tabular}
%    \vspace{2mm}
%    \caption{P-values that were found to be statistically significant ($p < 0.05$) from the one sided Mann-Whitney U test considering the difference in oxygen retention for a low (1 TO) and high (9, 10, or 11 TO) initial water content case. The middle column tests if the oxygen retention was worse at low initial water (1 TO) while the right column tests if the oxygen retention was better at low initial water (1 TO). A missing value indicates the hypothesis was not found to be statistically significant.}
%    \label{tab:MannWhitney}
%\end{table} 

\subsection{Comparison to Atmosphere-Interior Evolution Model} \label{subsec:JoshKTCompare}

Here we compare our calculated escape rates to \citet{krissansen2022predictions}, a similar study modeling the atmospheric evolution of the TRAPPIST-1 planets, but with a particular focus on the interior evolution and surface-atmosphere exchange. \citet{krissansen2022predictions} use the Planetary Atmosphere, Crust, and MANtle (PACMAN) evolutionary model to simulate terrestrial planet mantle and atmosphere evolution from an initial magma ocean phase through solidification to temperate geochemical cycling on the TRAPPIST-1 planets \citep[validated in][]{Krissansen2021venuseverhabitable,Krissansen2021oxygenfalsepos}. They complete 5000 model runs total for each planet; unlike this study, their simulations sample initial water in log space from 0.7 -- 300 TO. For the interior planets, T1-b, c, and d, they show half of all model runs produce $>$1 bar oxygen atmospheres while the other half result in airless bodies. Alike to the current study, oxygen-rich modern atmospheres become less probable as distance from the star increases, noting that while initial water has the largest effect on oxygen build-up for the inner planets, atmospheric escape physics and geological factors influence the outer planets more heavily. Overall, \citet{krissansen2022predictions} show that complete atmospheric erosion or oxygen-rich atmospheres are most likely for the inner planets, while the outer planets are more likely to possess CO$_{2}$ or CO$_{2}$/O$_{2}$ dominated atmospheres, the latter occurring if CO$_{2}$ efficiently acts as a secondary producer of oxygen through photodissociation.

\begin{figure}
    \centering
    \includegraphics[width=\textwidth]{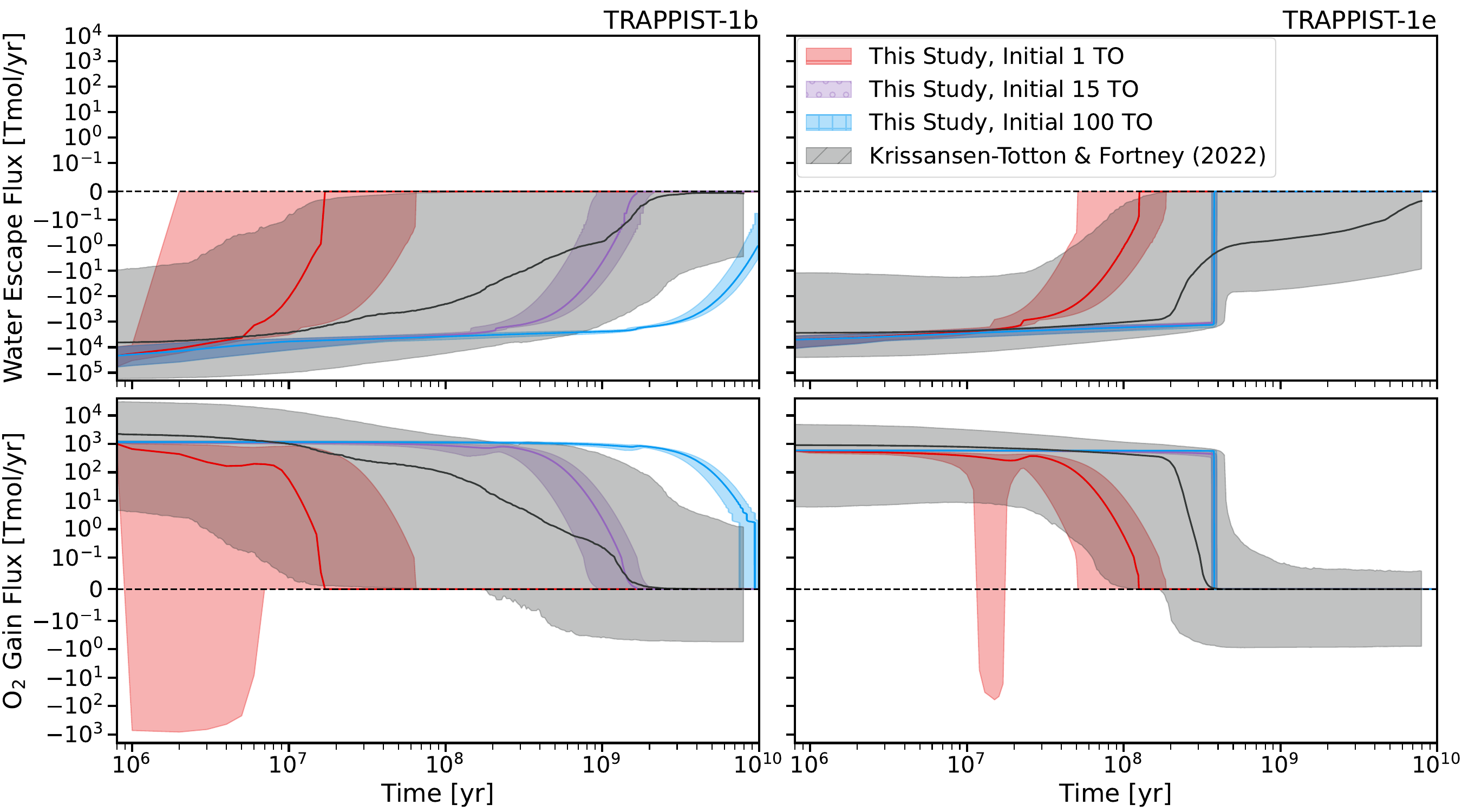}
    \caption{A comparison of the water loss and oxygen production rates from \citet{krissansen2022predictions} to this study. Design of the figure is meant to recreate the style published in \citet{krissansen2022predictions}; all shaded regions denote 2$\sigma$ uncertainties. The red, purple, and blue lines and shaded regions show the results from this study for initial water contents of 1, 15, and 100 TO, respectively, and the grey line and shaded region shows the results presented in \citet{krissansen2022predictions}. \textbf{Left Column:} Comparisons for the interior planet TRAPPIST-1 b \citep[see Figure 2 in][]{krissansen2022predictions}. \textbf{Right Column:} Comparisons for the habitable zone planet TRAPPIST-1 e \citep[see Figure 3 in][]{krissansen2022predictions}. \textbf{Top Row:} Rate of water loss due to thermal atmospheric escape [Tmol/yr] over time in years. Note, as water is being lost, the flux is negative. \textbf{Bottom Row:} Rates of oxygen production (i.e., gain flux) [Tmol/yr] due to water photolysis and subsequent hydrogen escape over time in years. In general, our results are in agreement with those of \citet{krissansen2022predictions}, and our median test case with an initial water inventory of 15 TO (log space average between 0.7 and 300 TO) stays within their model uncertainties for the entire age of the system. \PeerReview{Note, in the T1-e comparisons, our 15 and 100 TO cases display nearly identical evolution and overlap on the plot.}}
    \label{fig:KTCompare}
\end{figure}

Figure \ref{fig:KTCompare} shows a comparison of the rates of H$_{2}$O loss and oxygen build-up with 2$\sigma$ uncertainty envelopes, as presented in \citet{krissansen2022predictions}, with our work for T-1b, an interior planet, and T-1 e, a HZ planet. Both studies adopt the stellar evolution models of \citet{Baraffe2015stellartracks}, and XUV properties from \citet{birky2021improved}, meaning differences in our work are not due to dissimilar XUV environments. Since we consider initial water contents on a grid, rather than the randomly sampled input of  \citet{krissansen2022predictions}, we compare 3 of our initial water contents in Figure \ref{fig:KTCompare}; we show initial contents of 1 and 100 TO to bracket the median water contents from \citet{krissansen2022predictions}, and an initial content of 15 TO is shown as the average value of 0.7 and 300 TO in log space.

The agreement between our studies is favorable, with our median 15 TO case remaining within the 2$\sigma$ uncertainty region of \citet{krissansen2022predictions}. The average rates from \citet{krissansen2022predictions} and from all shown initial water contents in this study begin in near perfect agreement ($\lesssim$10$^{7}$ years), with our 1 TO case deviating first as desiccation is reached quickly. For T1-b, our 100 TO case begins to deviate more at $\sim$1 Gyr, showing relatively larger fluxes of both water escape and O$_{2}$ production as more water is available to sustain escape than the majority of model runs from \citet{krissansen2022predictions}. Our 15 TO T1-b case remains in close agreement for the entirety of the system's lifetime, although the shape of the curves show that escape rates in \citet{krissansen2022predictions} slow more gradually.  For T1-e, entrance to the HZ strictly ends evolution for our 15 and 100 TO cases at $\sim$380 Myr. However,  as \citet{krissansen2022predictions} did not use as stringent of a simulation halt, they still report minor water loss and oxygen production past this point, although it is significantly reduced. Before HZ entrance, our 15 and 100 TO T1-e cases behave similarly, but the 15 TO case does begin to deviate and show more quickly lowering rates suggesting it would remain in better agreement with \citet{krissansen2022predictions}, as expected. In general, 2$\sigma$ uncertainties on the results shown in this study are smaller because the curves each have a fixed initial water content, while random sampling of initial water content in \citet{krissansen2022predictions} contributed to larger uncertainty in simulation behavior.

\PeerReview{When oxygen drag is efficient, a ``negative gain flux" may occur, where oxygen is lost faster than it is photolytically produced, which could deplete existing atmospheric oxygen inventories. Figure \ref{fig:KTCompare} shows these negative O$_{2}$ gain fluxes early in the age of the system ($\lesssim$200 Myr), which we were only able to reproduce with VPLanet after the model modifications implemented here. We found that closer-in planets with lower initial water contents are more likely to experience a negative O$_{2}$ gain flux. For example, with an initial water content of 1 TO for T1-b and T1-e, 62\% of the T1-b and 35\% of the T1-e simulations experienced at least one time step with a negative O$_{2}$ gain flux, respectively; this number dropped to 16\% and 4\%, respectively, when initial water was increased to 2 TO.}

\PeerReview{Negative O$_{2}$ gain fluxes occur when the mixing ratio of oxygen is as high as possible without pushing escape into the diffusion-limited regime, and must occur during periods of high incident XUV (e.g., $\lesssim$200 Myr). As the mixing ratio of oxygen increases, loss to hydrodynamic drag also increases (e.g., Eq. \ref{eq:etao}); but if the mixing ratio of O$_{2}$ exceeds that of water, the atmosphere will enter diffusion-limited escape and oxygen drag will be shut-off altogether. Negative O$_{2}$ gain flux is therefore most likely to occur during periods when the incident XUV is high (e.g., $\lesssim$200 Myr), and when the mixing ratio of oxygen approaches the maximum value needed to trigger diffusion limited escape and the shut down of oxygen drag. A negative O$_{2}$ gain flux will then continue until diffusion-limited escape starts or the XUV flux has decreased sufficiently. Considering our 1 TO cases in Figure \ref{fig:KTCompare}, T1-b starts to lose net atmospheric O$_{2}$ almost immediately due to its close proximity to the star, whereas T1-e needs at least 10 Myr to enter a negative O$_{2}$ gain flux state, as some time is needed for the oxygen mixing ratio to increase sufficiently.}

\section{Discussion} \label{sec:Discussion}

We have used the most recently available constraints on the TRAPPIST-1 system \citep{agol2021refining,birky2021improved} and a hydrodynamic atmospheric escape model to statistically explore the ranges of possible water loss and oxygen buildup on the TRAPPIST-1 planets. Improving on past studies, we combine observational uncertainties on the stellar luminosity evolution and planetary parameters to construct self-consistent input chains that are randomly sampled for each simulation, fully incorporating the statistically plausible region of initial conditions for physical system parameters. Without considering outgassing, we find the interior planets (T1-b, c, and d) nearly, or completely, lose all of their initial surface water for all but the largest initial contents \PeerReview{(when initial water is $>$50 -- 100 TO)}, while the outer planets (T1-e, f, g, and h) only become desiccated for the very lowest initial water contents considered (1 -- 2 TO). Given that our model halts  hydrodynamic escape when a planet enters the habitable zone, the outer planets never lose more than 10 TO total, which results in maximum possible oxygen retention values spanning $\sim$90 -- 1290 bars. Comparing our results on  planetary water loss and oxygen production rates to those reported in \citet{krissansen2022predictions}, we find that rates from an example set of 3 of our tested initial water contents remained within their reported uncertainties for most, or the entirety, of the age of the system, showing favorable agreement.  

We find that atmospheric oxygen retention -- what fraction of the oxygen produced by water photolysis is retained in the atmosphere -- is a function of both planet-star separation and initial water inventory. Closer to the host star, we find that larger percentages of photolysis-produced oxygen are lost due to hydrodynamic drag, offsetting the substantial oxygen produced by water photolysis. For the cumulative water loss at low initial contents, our simulations show that the interior planets (T1-b, c, and d) retain $\sim$10\% -- 30\% of the oxygen generated, and the outer planets (T1-e, f, and g) retain $\sim$50\% -- 70\%. At high initial water contents, this retention efficiency approaches $\sim$70\% -- 90\% for all planets in the system, excluding T1-h. However, despite the enhanced loss rate of oxygen, for initial water contents of more than 4-5 TO, the overall production rate is sufficiently high that the interior planets always retain more O$_{2}$ than the outer planets. Consequently, for initial water contents $\lesssim$4 TO, the outer planets are more likely to retain atmospheric oxygen. 

We retrieved the suite of plausible water evolutionary histories consistent with a known or predicted final water mass fraction (WMF) for the outer planets, demonstrating a mechanism to determine the most probable initial water content of the planet for a predicted final water content. For the maximum present-day WMFs constrained by measured planetary densities \citet{agol2021refining}, we find the most probable initial water contents for all outer planets to be in the range of $\sim$80 -- 300 TO. For T1-e, we interpolated from this probability data to report the 1$\sigma$ confidence interval on our maximum initial water content prediction of 90.6$_{-37.2}^{+70.6}$ TO, based on the model we employ and the present-day water constraints we adopt. This analysis is a proof of concept for a method to retrieve probable initial water content from an assumed present-day state. However, our model assumptions will produce upper limits, and the true initial water content may be much lower, as we assume that 1) the planets' low densities are entirely due to their WMF, and 2) that water was not sequestered in the planetary interior, but was always on the surface and accessible to atmospheric loss. \PeerReview{To elaborate on the latter point, if water was instead partially or fully trapped in the planetary interior, less initial water would be required to recreate the same WMF at present-day, as it would be protected from loss due to escape.}

%ncluding the nature of the planets' low densities and the lack of geological processes in our escape treatment 

%In this section, we discuss the implications of our results, trends in the data, and the limitations and assumptions made in our work. We begin by discussing the behavior of water loss and how it is affected by each parameter considered (\S \ref{subsec:WaterlossDiscuss}), followed by the trends and behaviors governing oxygen retention efficiency (\S \ref{subsec:OxygenDiscuss}), and how these results compare to other work (\S \ref{subsec:ComparisonDiscuss}). We then discuss the broader significance and implications of our study (\S \ref{subsec:ImportanceDiscuss}), including the benefits of retrieving water evolutionary history (\S \ref{subsec:PredictingInitialDiscuss}) and how this study can help us interpret the first JWST observations of T1-b (\S \ref{subsubsec:JWSTObsDiscuss}). We then transition to considering the big picture of planetary environments to understand how our results may be affected by the inclusion of planetary processes not considered here and approximations made within our model (\S \ref{subsec:LimitationsDiscuss}, and subsections therein).

\subsection{Behavior of Water Loss} \label{subsec:WaterlossDiscuss}

%Our simulations show that factors including time spent outside or inside the habitable zone, distance from the host star, and initial water content affect both the instantaneous rates of water loss, and the end state observed. The cumulative water loss is highly dependent on time spent in the hydrodynamic regime, and consequently incident XUV flux and initial water inventory. 
Our work shows that water loss rates and cumulative water lost are highly dependent on the incident XUV flux and the atmospheric mixing ratio of water; implying both orbital distance and initial water content exert strong influence on the end state observed. Qualitatively, increasing orbital distance will decrease water loss by decreasing incident XUV, and decreasing the time to HZ entry and hydrodynamic escape shutoff.  Conversely, increasing initial water content will increase water loss rates by creating a higher atmospheric water mixing ratio at earlier ages during greater incident XUV. Across the conditions considered in this work, the interior planets (T1-b, c and d) frequently lose their entire initial water inventory, while the outer planets (T1-e, f, g and h) are more likely to retain water to the present-day. 

As the interior planets are located close enough to the star to experience escape for the entirety of the system's lifetime, their end state is most heavily influenced by initial water content (see Figure \ref{fig:EscapePathways}), which is in agreement with \citep{krissansen2022predictions}. Conversely, the end state behavior of the HZ or outer planets (T1-e, f, g, and h) is primarily influenced by distance from the host star, which affects total integrated XUV flux experienced by the planet, and also determines the time to habitable zone entrance (i.e., total time spent in the hydrodynamic escape regime). 
%and thus total integrated XUV flux experienced by the planet.
For the outer planets, escape is shut off at HZ entrance due to the assumed formation of a cold trap; however, we caution that the true efficacy of a cold trap to prevent hydrodynamic escape may not be this efficient, and may be influenced by other factors (see also \S \ref{subsec:LimitationsDiscuss}). As HZ entrance results in a short period of escape ($\sim$50 -- 400 Myr) relative to the age of the system ($\tau\sim$8 Gyr) for the outer planets, desiccation is rarely witnessed and the maximum amount of water they may lose through hydrodynamic escape is $<$10 TO in all cases in this study, and as low as 0.8 TO for T1-h. \PeerReview{Though the outer planets may better constrain initial water content when considering measurements of present-day WMF, measuring  the present-day WMF is observationally challenging as water may be sequestered in the interior or in a surface ocean, and so not amenable to transmission observations, and also difficult to derive uniquely from planetary density constraints \citep{agol2021refining}. However, initial water more heavily influences the present-day atmospheric oxygen abundance on the interior planets, and ocean-loss generated O$_{2}$, or its proxy O$_{3}$, may be accessible in JWST observations \citep[e.g.,][]{lustig2019detectability} and next generation ground-based telescopes \citep[e.g.,][]{Currie2023ELT}, and so may provide the first strong constraints on the initial water content of the planetary system as a whole, when considering current and near-term observational capabilities.} 

%%%% SUGGEST YOU BRING COMPARISON TO PREVIOUS WORK FOR WATER LOSS UP HERE.  IT CAN BE A SUBSECTION. 
\subsubsection{Comparison with Previous Work} \label{sec:prevworkwater}

Comparison with past studies shows agreement in our predictions that the inner planets are more likely to be desiccated, but the outer planets retain water, and that stellar XUV and activity evolution have larger impacts on total water loss than geological processes.  Our results confirm numerous past studies that desiccation is possible on the interior planets of TRAPPIST-1, or a similar M dwarf system, for all but extremely large initial water contents \citep{LugerBarnes2015extreme,schaefer2016predictions,bolmont2017water,bourrier2017temporal,Johnstone2020hydrodynamicesc,krissansen2022predictions}.  Similarly, we also find maximum water loss on the HZ and outer planets of $\sim$1 -- 10 TO, implying these planets may still possess water at present-day \citep{LugerBarnes2015extreme,bolmont2017water,bourrier2017temporal,barth2021magma,krissansen2022predictions}. Comparison with past studies shows that predicted water loss is very sensitive to assumptions on the stellar XUV and activity evolution, implying that tighter constraints on TRAPPIST-1's stellar spectrum and activity can greatly improve predictions for planetary outcomes. For example, \citet{bolmont2017water} (considering T-1b and c) and \citet{bourrier2017temporal} (considering all 7 planets in the system) adopt XUV saturation fractions $\sim$1.32x and $\sim$2.5 -- 3.5x smaller than this work, respectively. Consequently, water loss predictions for the inner planets in this study are up to 3x larger than \citet{bolmont2017water} and $>$3x larger than \citet{bourrier2017temporal}. Comparison of our results, which do not include geological processes, with those of \citet{krissansen2022predictions}, that do, suggest that the inclusion of geologic processes (e.g., sequestration of water in the planetary interior and outgassing over time) has less of an influence on total water loss than changing XUV environment, since \citet{krissansen2022predictions} adopts the same XUV parameterizations \citep{birky2021improved} as this study and reports agreement on overall conclusions on the final planetary water contents. These comparisons highlight how marginal increases in XUV flux may produce drastic increases in water loss, and suggest more observations of the TRAPPIST-1 stellar environment may translate to significant improvements in model predictions.

\subsection{Oxygen Production, Retention and Loss} \label{subsec:OxygenDiscuss}

%Despite its role in life on Earth, the strength of oxygen as a biosignature on the TRAPPIST-1 planets is questionable \citep[e.g.,][]{Wordsworth2014oxygenDomAtmHZ,LugerBarnes2015extreme,meadows2018o2InContext}, and 
The upper limits on oxygen production and retention quantified by this study provide important environmental context on atmospheric O$_{2}$ in the TRAPPIST-1 system, and while retained atmospheres less than 1 bar may constitute a false positive for an \ce{O2} biosignature \citep{schaefer2016predictions,meadows2017reflections}, significantly larger amounts may be a clear indication of an abiotic post-ocean loss atmosphere \citep{LugerBarnes2015extreme}.  
%NEED SENTENCE OR TWO HERE ON YOUR MAJOR OXYGEN PRODUCTION TAKEAWAYS, AND THE REMINDER THAT IT IS ONE THING TO PRODUCE IT, BUT THAT YOU HAVE TO KEEP IT, AND THAT YOU CONSIDERED AND FOUND OXYGEN RETENTION TO BE SENSITIVE TO A NUMBER OF PROCESSES...
On Earth-sized terrestrials, the loss of an ocean's worth of water will generate $\sim$250 bars of O$_2$, and the inner planets, which lose more water over the system lifetime, can produce hundreds to thousands of bars atmospheric oxygen, depending on initial water content. However, not all this oxygen is retained, and significant losses can occur due to energetic hydrogen escape flows that remove the oxygen from the atmosphere. We find that cumulative oxygen retention increases with increasing distance from the star; however, for initial water contents greater than 4 -- 5 TO, the interior planets retain more oxygen cumulatively than the outer planets due to greater overall water loss (and the subsequent increase in oxygen production). 

Similarly, we find the efficiency of oxygen retention increases with increasing distance from the star and/or increasing initial water content (\S \ref{subsec:O2Efficiency}, Figure \ref{fig:RetentionEfficiency}). The distance dependence is due to greater XUV flux at small orbital separations, resulting in more energetic escape flows and oxygen loss through drag; however, planets with low gravity, including T1-d and h, may deviate from this pattern and lose more oxygen than their closer-in neighbors. The increase in the efficiency of oxygen retention with increasing initial water content  is driven by larger water inventories producing a large atmospheric water to oxygen ratio, which decreases the rate of oxygen loss through drag.

Low oxygen retention efficiencies are also more likely to occur when oxygen mixing ratios are higher, but before the oxygen fraction exceeds the water fraction in the atmosphere, triggering diffusion limited escape and oxygen retention.  Extremely low initial water contents can lead to high oxygen mixing ratios early in the system's lifetime, during the greatest incident XUV, resulting in highly efficient oxygen drag and lowered retention efficiency -- in some cases leading to total oxygen loss and atmospheric erosion for the interior planets. A similarly lowered retention efficiency could also occur 
%since retention efficiency's dependence on initial water  is due to the resulting atmospheric mixing ratios, 
if the initial planetary water inventory was high, but sequestered in the interior and slowly outgassed, rather than being continuously available for escape at the surface as is often assumed \citep[e.g.,][]{krissansen2022predictions}.  %then this would 
%water trapping and outgassing from the interior would serve to also lower the retention efficiency, 
In this case, the throttling of the water availability by outgassing would be more likely to result in lower atmospheric water mixing ratios at any given time leading to greater oxygen loss through drag. 

\subsubsection{Comparison with Previous Work}

In contrast to the general agreement on water loss predictions we see across previous studies (\S\ref{sec:prevworkwater}), predictions of oxygen retention for M dwarf planets vary. In this work, we witness a broad range of oxygen retention scenarios depending primarily on initial water content and orbital distance. For medium to high initial water inventories (typically above $\sim$5 TO) we predict comparatively high planetary oxygen retention (e.g., hundreds to thousands of bars), which is in agreement with several previous studies \citep{LugerBarnes2015extreme,bolmont2017water,bourrier2017temporal,krissansen2022predictions}. In contrast, others show that little to no oxygen would survive hydrodynamic escape processes (e.g. zero to tens of bars) even at medium to high initial water contents \citep{schaefer2016predictions,Johnstone2020hydrodynamicesc}.
%-- with some studies, including this work, predicting comparatively high planetary oxygen retention (e.g. hundreds to thousands of bars) \citep{LugerBarnes2015extreme,bolmont2017water,bourrier2017temporal,krissansen2022predictions}, and others showing that little to no oxygen would survive hydrodynamic escape processes (e.g. zero to tens of bars)  \citep{schaefer2016predictions,Johnstone2020hydrodynamicesc}.
\PeerReview{Within these predictions, there is further disagreement as to the precise quantity of oxygen retained, and if it would remain in the atmosphere \citep[e.g.,][]{LugerBarnes2015extreme} or be removed partially or entirely by reactions with iron in the planet's mantle \citep[e.g.,][]{Wordsworth2018redox,barth2021magma,krissansen2022predictions}.}

At low water contents ($\lesssim$4 TO), this work indicates 
%This work (Figure XX?) and several previous studies have indicated 
that large percentages, or even the entirety of photolytically produced oxygen may be lost solely through drag during hydrodynamic escape (as demonstrated in Figures \ref{fig:StatPlotHZ}, \ref{fig:RetentionEfficiency}, and \ref{fig:KTCompare}); similar results with complete oxygen loss have been reported by several previous studies \citep{schaefer2016predictions,Johnstone2020hydrodynamicesc,krissansen2022predictions}. \citet{schaefer2016predictions} considers water contents up to 1000 TO in the GJ 1132 system \citep[M$_{*}$ = 0.181 M$_{\bigodot}$, see][]{BertaThompson2015gj1132b} and show that the super-Earth GJ 1132 b would likely lose 90 -- 99\% of all photolytically produced oxygen to space through drag. \PeerReview{Likewise, the \citet{schaefer2016predictions} model was also adopted by \citet{Kreidberg2019lhs3844b} which reported that initial water contents $<$240 TO resulted in complete atmospheric loss on LHS 3488 b. Comparing GJ 1132 b and LHS 3488 b to T1-b in the TRAPPIST-1 system, they have substantially larger insolation -- with GJ 1132 b receiving about 19 times more flux than Earth \citep{schaefer2016predictions} and LHS 3488 b receiving 70 times more flux than Earth \citep{Vanderspek2019lhs3488bdiscov}, whereas T1-b receives only about 4 times more than Earth \citep{agol2021refining}.} This higher incident radiation helps to explain why such high oxygen loss rates are reported. Similar oxygen behavior was shown by \citet{Johnstone2020hydrodynamicesc} when considering an Earth-like planet with an M dwarf host, reporting water loss up to 120 TO during the stellar saturation time (the first 1 Gyr) with little to no planetary oxygen retention as a result of efficient drag. However, when considering medium to high initial water contents ($>$4 -- 5 TO) and also accounting for drag, our results instead predict a significant build-up of oxygen. This contradicts the work of \citet{Johnstone2020hydrodynamicesc}, but is consistent with several other studies, which predict oxygen build-up ranging from several hundreds \citep{bolmont2017water} to thousands of bars \citep{LugerBarnes2015extreme} on the interior planets, and several tens \citep{bolmont2017water} to hundreds of bars \citep{LugerBarnes2015extreme,barth2021magma} on the outer planets. These differences in predictions are likely because, as we have shown, the precise amount of build-up is highly dependent on duration of escape, planetary gravity, and mathematical descriptions of crossover mass and oxygen flux. Observations to constrain atmospheric oxygen on the TRAPPIST-1 planets would help constrain evolutionary scenarios and outcomes, and improve theoretical models.

\subsection{Constraining the Early Water Abundance of the TRAPPIST-1 Planets}

We explore the process of constraining initial water content of the TRAPPIST-1 planets, both through model inference (\S\ref{subsec:InitialWaterPredict}) and by considering recent JWST measurements. By combining simulations and JWST data from multiple planets -- in this case, TRAPPIST-1 b and c -- we can derive much tighter constraints on initial surface water abundance (\S \ref{subsubsec:JWSTpredictresults}).
%than is possible with our current modeling capability
%We explore through a comparison of JWST observations of TRAPPIST-1b and c (\S 
%\ref{subsubsec:JWSTObsDiscuss}).

\subsubsection{Statistically Retrieving Planetary Initial Water Content} \label{subsec:PredictingInitialDiscuss}

As we show in Section \ref{subsec:InitialWaterPredict}, given present-day water and oxygen inventories, our model suite of escape histories can provide statistical constraints on  initial water content and past water loss.  While using constraints on present day water inventories informed by measured planetary densities can provide a key constraint \citep{agol2021refining}, future observational evidence on the presence and nature of the TRAPPIST-1 atmospheres could be used to refine initial water content and subsequent planetary evolution.  For example, if an outer planet is found to be desiccated, it would place a strict upper limit on initial water content, given the much shorter escape lifetimes for outer planets. Correspondingly, desiccation on an interior planet provides weaker constraints as there  is a much broader range of initial water abundances that could still result in desiccation over the system lifetime (\S\ref{subsubsec:JWSTObsDiscuss}). For a desiccated interior planet, oxygen may be a better indicator of ocean loss and other evolutionary processes as it is much more tightly correlated to initial water content (see Figure \ref{fig:StatPlotInterior}). If a planet is observed to still possess water, the method explored in Section \ref{subsec:InitialWaterPredict} and Appendix \ref{appendix:initialpredictionMethods} demonstrates a basic framework for conducting statistical inference to achieve an initial upper limit constraint on initial water.

%The process of inferring system evolution from observations, which we also refer to as ``retrieving evolutionary history'', will become increasingly relevant to the astrobiological community as we continue to improve our capability to observe terrestrial exoplanet atmospheres. 
%The case explored here lays the groundwork for conducting this evolutionary history retrieval in a statistically significant way, but this basic inference becomes more untenable as we aim to fit higher dimensional models and combine observations of different variables to teach us about the overall planetary environment. Thus, these methods must evolve in tandem with our evolutionary models and observation efforts. 
Statistical inference may also be leveraged as a model intercomparison tool, and to provide critical constraints on planet formation, migration and water delivery models. In the example of the current work, if multiple studies on water loss conduct an analysis to map initial to final water content, we can indirectly compare the latent space (e.g., exact equations, order of operations, timestepping, etc) predicted by those models during evolution. In a more concrete sense, our method of retrieving initial from final water content applied to the TRAPPIST-1 system can provide model constraints for formation and water delivery scenarios of the system. Even with relatively uncertain constraints on initial water content, this may inform if the planets formed \textit{in situ} or past the snow line followed by inwards migration \citep{Raymond2022upperlimT1}.

\subsubsection{Combining JWST Observations of T1-b \& T1-c to Further Constrain Initial Water Inventory} \label{subsubsec:JWSTObsDiscuss}

\PeerReview{Recent observations of TRAPPIST-1b and c have demonstrated that the former may be an airless world \citep{Greene2023T1bJWST,Ih2023t1binterpret}, while the latter may be better fit by a tenuous oxygen-dominated atmosphere \citep{Zieba2023T1cJWST,Lincowski2023T1c}. In Section \ref{subsubsec:JWSTpredictresults}, we used MCMC to fit for the initial water content that would produce this final atmospheric state when assuming basic oxygen sink estimations and that the two planets formed with comparable water inventories; the results of this analysis are shown in Figure \ref{fig:JWSTAnalysis}. Our MCMC reported that the minimum initial water content of the planets is 8.2$^{+1.5}_{-1.0}$ TO. This estimated range is in agreement with \citet{Zieba2023T1cJWST}, who suggest an initial water mass of 9.5$^{+7.5}_{-2.3}$ TO through an analysis of conservative CO$_{2}$ upper limit constraints, albeit with a similar atmospheric escape model \citep{LugerBarnes2015extreme,schaefer2016predictions}.}

These constraints on minimum initial water using JWST observations of T1-b and c, and our estimates of subsequent water and oxygen escape and loss, have important ramifications for the presence of water, and habitable conditions, on the TRAPPIST-1 habitable zone planets. If all planets in the system formed with comparable initial water contents, as supported by the delicate resonance chain and recent formation studies \citep[e.g.,][]{Raymond2022upperlimT1}, our simulations suggest that an initial water content of \PeerReview{8 TO} would imply the HZ and outer planets would still possess water after escape is shut off upon entrance to the habitable zone. \PeerReview{With an initial water content of 8 TO, we find T1-e, f, g, and h would possess 1.7$^{+0.47}_{-0.57}$, 3.6$^{+0.32}_{-0.46}$, 4.8$^{+0.21}_{-0.29}$, and 7.2$^{+0.14}_{-0.20}$ TO, respectively, after HZ entrance.} Thus, a confirmation of oxygen in the atmosphere of T1-c would increase the probability that the HZ planets retained water and could be potentially habitable. 
%strongly suggest the existence of water on the HZ planets, increasing the probability that they are in fact habitable.

If T1-c is airless or nearly airless, \PeerReview{this still may provide information on an upper limit on initial water of the system, for example }this would narrow our previous upper limit constraint from T1-b alone; following the same removal estimates described in section \ref{subsubsec:JWSTpredictresults}, we find an airless T1-c would suggest an upper limit on initial surface water of 9 TO. Alternatively, in this airless T1-c scenario, \citet{Zieba2023T1cJWST} predict initial water content would be reduced to $\sim$4 TO; for this initial water, our model predicts T1-e, f, g, and h would possess 0.03$^{+0.05}_{-0.03}$, 0.52$^{+0.15}_{-0.17}$, 1.2$^{+0.14}_{-0.19}$, and 3.22$^{+0.14}_{-0.20}$ TO, respectively, after HZ entrance. Although this case would imply a greater risk of water loss on T1-e and f, it would not rule out surface water on these planets and would still support considerable endowments on T1-g and h. This is in agreement with the conclusion of \citet{Krissansen2023nondetectJWST} that an airless T1-b (and c) does not preclude the outer planets from possessing substantial atmospheres.

%would not require the outer planets to be desiccated.

% Additionally, geologic processes and photochemical shielding may reduce water loss during the escape period, resulting in larger present-day water endowments than those suggested here; though, planetary evolution past the time of HZ entrance may result in further water loss, particularly if a super-heated atmosphere enabled hydrodynamic escape to continue past HZ entrance. 

If the TRAPPIST-1 HZ and outer planets are likely to possess present-day oceans, as the recent JWST observations \citep{Greene2023T1bJWST,Zieba2023T1cJWST} interpreted with our models suggest, then these will make important  targets for JWST observations. When observing these targets in transmission, however, the further work will be required to remove the effects of stellar contamination from the data, as recent observations have shown to be prevalent \citep{Rackham2018TLE,Lim2023T1Flare}. 
%in the context of our water loss distributions indicate the HZ and outer planets of the TRAPPIST-1 system may still possess oceans at present-day, motivating follow-up observations of these planets. 
%In considering planetary oxygenation, at an initial content of 9 TO, T1-b produces more oxygen overall and displays better oxygen retention efficiency through the hydrodynamic loss process ($\sim$54\%) than T1-d ($\sim$49\%) and T1-h ($\sim$46\%) (see Figure \ref{fig:RetentionEfficiency}), suggesting these planets will very likely be O$_{2}$-poor if they experience similar geologic and non-thermal escape oxygen sinks as T1-b.
Though O$_{2}$ has been shown to be extremely difficult to detect with JWST, particularly for the outer planets, which will be restricted to transmission measurements \citep[e.g.,][]{lustig2019detectability,wunderlich2019detectability,gialluca2021characterizing,Gialluca2023corrigendum,Meadows2023jwst}, post-ocean loss O$_{2}$ atmospheres may be detectable on the interior planets by using secondary eclipse MIRI observations to search for the O$_3$ that is produced by O$_{2}$ photolysis.

%In particular, the mid-infrared ozone band with the JWST MIRI instrument \citep{lustig2019detectability}, and the interior planets are hot enough to be observed with secondary eclipse photometry \citep[as demonstrated by][]{Greene2023T1bJWST,Zieba2023T1cJWST}. Thus, their high final state dependence on initial water content and high susceptibility to post-ocean loss O$_{2}$-dominated atmospheres, which agree with current data \citep{Lincowski2023T1c}, suggest measuring O$_{3}$ as a proxy for O$_{2}$ on the interior planets will provide a plethora of information on the evolutionary history of the system as a whole.

\subsection{Broader Planetary Environment \& Model Limitations} \label{subsec:LimitationsDiscuss}

Future changes to our understanding of initial conditions, model treatment of escape, and the planetary environment post-hydrodynamic loss may influence our results, and we discuss potential ramifications here. Several changes within the model itself may create a more realistic treatment of the escape process, most notable of which would be the addition of a defined atmospheric vertical structure and variable pressure-temperature profile, and the inclusion of non-thermal escape; these beneficial model changes are discussed in Section \ref{subsubsec:modeldiscuss}. In the current study we have explored a broad range of initial water contents at a fixed escape onset time of 1 Myr; however, formation and water delivery research on the TRAPPIST-1 system may provide valuable information that would allow us to narrow the allowed range of initial water contents and conserve computational expense -- these considerations are discussed in Section \ref{subsubsec:watercontentdiscuss}. All of these considerations should be kept in mind when using this work to interpret observations of the system. 

\subsubsection{Beneficial Model Changes \& Additions} \label{subsubsec:modeldiscuss}

Several model updates would make our treatment of the hydrodynamic escape period more realistic, and we find the overwhelming majority of which would likely serve to reduce the cumulative water loss, suggesting the results presented in this study are upper limits to water loss and oxygen accumulation. Since our simulations currently assume atmospheres with perfect mixing and constant pressure and temperature, approximations in escape physics are required as vertical transport and temperature gradients are neglected. Approximations commonly made in escape modeling \citep{LugerBarnes2015extreme,tian2015atmospheric,schaefer2016predictions,bourrier2017temporal} may over- or under-estimate loss rates, particularly in the energy-limited formalism \citep{Krenn2021energylimapproxassess} or at boundary conditions between escape regime changes \citep{Volkov2011thermalescapetransition,gronoff2020atmospheric}. Our description of hydrodynamic escape follows the energy-limited approximation \citep{hunten1973escape,Watson1981escformalism} however, it has been shown that this formalism may considerably deviate from hydrodynamic mass loss models in certain scenarios. Based on \citet{Krenn2021energylimapproxassess}, which assesses the scenarios in which the energy-limited approximation is applicable, all the planets in our study remain in agreement with the more rigorous hydrodynamic loss simulations of \citet{Krenn2021energylimapproxassess} within a factor of 10 for the entirety of our simulations, except T1-d, and h, and T1-e at the end of their simulations. 
%These three planets move closer to disagreement with \citet{Krenn2021energylimapproxassess} greater than a factor of 10 at the end of their simulations due to decreasing XUV flux, with T1-d and h more likely to deviate than T1-e. 
%Furthermore, while this work adopted atmospheric compositions of pure water and its constituent atoms, the inclusion of species beyond water will affect escape, production, and heating or cooling rates. 

\PeerReview{Other escape processes not considered in this work may also be acting upon the atmosphere, such as non-thermal escape \citep[e.g.,][]{Dong2018atmescT1}, photochemical escape \citep[e.g.,][]{Wordsworth2018redox}, and core-powered mass-loss \citep[e.g.,][]{Ginzburg2016corepowered,Gupta2019corepowered}. Though in the case of core-powered mass-loss, this is more typically explored for super-Earth and sub-Neptune populations, and will likely have a greater effect on an initial H/He envelope \citep{Gupta2019corepowered} which we did not take into consideration in this work.}

In this work, no photochemistry besides photolysis, and no geological evolution or surface-atmosphere exchange was modeled. In reality, photochemical reactions play a pivotal role in atmospheric escape, particularly in UV shielding, the creation of photochemical products that may act as UV shields, and effects on atmospheric heating and cooling. Through UV shielding, particles at higher altitudes may absorb UV radiation and protect molecules lower in the atmosphere from photodissociation, including self-shielding (e.g., O$_{2}$ protecting other O$_{2}$ molecules) and shielding between molecular species \citep{Calahan2022selfshield}. Of notable importance, the absorption cross sections of water, O$_{2}$, O$_{3}$, and CO$_{2}$ overlap heavily \citep{Parkinson2003watcrosssec,Domagal2014abioticO}, thus the intense build-up of O$_{2}$ we observe in our models may serve to protect water from photodissociation, lowering mass loss rates. Furthermore, the photochemical production of ozone (O$_{3}$) from free O$_{2}$ would be highly active on a planet with an over abundance of free oxygen \citep{Segura2005Mstarbiosignatures,meadows2018o2InContext}, increasing the amount of UV shielding molecules and greenhouse gases in the atmosphere. CO$_{2}$, prevalent on all the terrestrial planets with atmospheres in the Solar System and considered a likely constituent of any TRAPPIST-1 planetary atmospheres \citep{lustig2019detectability,lincowski2018Trappist,Lincowski2022eratum,krissansen2022predictions}, is also an important constituent to consider in super heated planetary atmospheres.  CO$_2$ may prolong escape through heating, or bottleneck escape through cooling effects in the middle and upper atmosphere \citep{Kulikov2007comparativess,Wordsworth2013watlosswithco2}. 
%and the production of CO$_{2}$ on the planets considered here is most likely prevalent .

Geological evolution, particularly a magma ocean phase, will be present and active during the hydrodynamic escape regime, with the attendant processes of water trapping in the mantle, outgassing, and the removal of oxygen or other species from the atmosphere  \citep{barth2021magma,krissansen2022predictions}. Hydrodynamic escape of a steam atmosphere is typically thought to be accompanied by a magma ocean phase \citep{Hamano2013typesmagmaoc,Acuna2021T1Hydrospheres}, as water provides a significant greenhouse effect that can prolong crustal solidification time.
%This magma ocean leads to appreciable surface-atmosphere exchange through water trapping and outgassing, oxygen removal, and the exchange of other molecules such as carbon-bearing species \citep{barth2021magma,krissansen2022predictions}.
A magma ocean may also prolong periods of escape by slowing the supply rate of water to the atmosphere, effectively reducing cumulative water loss. However, though cumulative water loss would be reduced, the atmosphere itself may be dry for larger initial water contents when including a magma ocean due to water trapping in the mantle \citep{barth2021magma}. A magma ocean would also suppress rapid build-up of atmospheric oxygen, prolonging energy-limited escape and increasing planetary oxygen retention by trapping it in the mantle. However, as initial water content increases, atmospheric oxygen build-up may still be inevitable, even with a magma ocean \citep{barth2021magma}. With a defined vertical structure (e.g., 1D atmospheric model), surface-atmosphere exchange during a magma ocean phase can be modeled explicitly, which would inform upwards transport of atmospheric constituents.

\subsubsection{Initializing \& Changing Surface Water Content} \label{subsubsec:watercontentdiscuss}

As initial water content represents our broadest initial parameter space, planetary formation and late water delivery research could narrow the range of plausible initial conditions used as input, providing tighter constraints on predicted outcomes for planets in the system, and also conserving computational expense, which could enable an increase in model complexity leading to more realistic treatment. Here we conservatively assume formation and water delivery is complete by 1 Myr.  However, if a longer formation and water delivery time casued escape to begin later when the incident XUV was lower, then cumulative water loss would be reduced, and the ratio of oxygen retained to water lost would be increased. %For example, the Solar System planets, which have much longer orbital periods and so likely longer formation times, are thought to have formed by approximately 100 Myr \citep{Albarede2009volatileaccretion}, with the bulk of water delivery peaking at $\sim$100 Myr \citep{Morbidelli2000cometsnodeliver,Raymond2004simsformation,Albarede2009volatileaccretion}. If the TRAPPIST-1 system reflected a similar timeline, there would be a noticeable decrease in cumulative water loss, but 
%, as the bulk of escape would occur in a decreased XUV environment compared to the first 100 Myr of the system. 
However, \citet{Raymond2022upperlimT1} show that the growth of the TRAPPIST-1 planets was likely complete within a few Myr with the bulk of their water reservoirs likely accreted during formation, as late water delivery greater than 5\% of 1 Earth mass would disrupt the orbital resonance we observe today. In this case, the timeline of our simulations is consistent with these model results, suggesting the planets would have been subjected to the earliest, and most intense, period of hydrodynamic blow-off.
%and as the idea of cometary water delivery during the late veneer \citep{Matsui1986veneer} has been contested by D/H ratio measurements \citep{Morbidelli2000cometsnodeliver,Raymond2004simsformation}, it is been suggested that the bulk of water delivery peaked at $\sim$100 Myr as well \citep{Albarede2009volatileaccretion}.

Regarding initial water content, previous work shows the TRAPPIST-1 planets may have formed with or possess up to $\sim$10 wt\% water \citep{Schoonenberg2019t1formation,agol2021refining,Acuna2021T1Hydrospheres,Raymond2022upperlimT1}, which would provide $\sim$430 TO on a 1 M$_{\bigoplus}$ planet, roughly consistent with the largest initial water inventories we considered. However, based on the recent T1-b and c observations \citet{Greene2023T1bJWST,Zieba2023T1cJWST}, our work would suggest the planets could not have formed with several hundreds of TO. The lower limit on initial water content, however, is relatively unconstrained by formation studies and extends to 0 for most planets \citep{Raymond2022upperlimT1}, which would remain plausible given the findings of this study, unless T1-c is later confirmed to possess atmospheric oxygen (see \S\ref{subsubsec:JWSTObsDiscuss}).  Overall, our timeline of water loss and initial water contents considered are within a plausible range given recent formation and water delivery scenarios, but computational expense may be conserved by exploring a narrowed range of initial water contents.

\section{Conclusions} \label{sec:Conclusion}

We have shown the plausible range of water loss and abiotic oxygen production on the TRAPPIST-1 planets due to hydrodynamic thermal atmospheric escape using an updated VPLanet model \citep{Barnes2020VPLanet} and the most recently available constraints on the TRAPPIST-1 system \citep{agol2021refining,birky2021improved}. We find the interior planets T1-b, c, and d, are likely desiccated for all but the largest initial water contents ($>$60, 50, \& 30 TO, respectively), and are at the greatest risk of complete atmospheric loss due to their proximity to the host star.  However, they may still retain significant oxygen at very large initial water contents depending on assumed sink processes. The outer planets likely retain some fraction of water for all but the lowest initial water contents ($\sim$1 TO). We find T1-e, f, g, and h lose, at most, approximately 8.0, 4.8, 3.4, and 0.8 TO, respectively, corresponding to maximum oxygen production values of approximately 1290, 800, 560, and 90 bars, respectively. Additionally, we have found oxygen retention efficiency may vary widely across distance and initial water content, with rates $<$10\% for close in planets with low initial water up to rates of $>$80\% for distant planets, and those with high initial water. 

%We further explored the process of retrieving initial water content and evolutionary history from a present-day water content, and demonstrated proof of concept for how to calculate this value in a statistically robust way. Through these predictions we conclude, if the present-day water mass fraction constraints from \citet{agol2021refining} are correct, the outer planets likely formed with 80 -- 100 TO. It should be noted, however, that this prediction would be in disagreement with those found while interpreting recent JWST observations of T1-b and c under the assumption that all planets in the system formed with comparable initial water contents.

%This work allows us to draw statistically robust conclusions on the nature of thermal escape across changing initial conditions, identifying plausible environments well-suited for complex modeling in follow-up work. The importance of changing XUV flux and the initial water content on escape outcomes is highlighted here, both affecting the rates of water loss and oxygen production, and the specific mechanisms of escape that are activated, such as the diffusion-limited regime. The process explored here is shown to have a first-order impact on habitability by governing the evolution and viability of surface water and by affecting a planets overall ability to sustain an atmosphere. As abiotic oxygen is a consequence of this process, we caution against relying on oxygen as a biosignature in late M dwarf environments without detailed environmental context. 

Using  currently available JWST observations for both T1-b and c \citep{Greene2023T1bJWST,Zieba2023T1cJWST}, our work constrains planetary initial water content, and suggests that the outer planets may still possess water at present-day. In the case of an airless T1-b, and a T1-c with a positive oxygen detection, an initial water content of \PeerReview{8.2$^{+1.5}_{-1.0}$} would be most likely from our study, in agreement with predictions of \citet{Zieba2023T1cJWST}. Assuming the outer planets are endowed with comparable initial water, an initial content of 8 TO implies T1-e, f, g, and h all likely possessed \PeerReview{$\gtrsim$1.5 TO} after HZ entrance, with the predicted amount increasing with orbital distance. If T1-c is found to be airless, we suggest a basic upper limit on initial surface water of 9 TO and \citet{Zieba2023T1cJWST} suggest an initial water content upper limit of $\sim$4 TO. Note, an initial water content of 4 TO does not definitively lead to desiccation on the outer planets in our work, which is in agreement with \citet{Krissansen2023nondetectJWST}. 
%As T1-b is likely to have better oxygen retention than T1-d or h, our work suggests these planets are highly likely to have oxygen poor atmospheres. 
These conclusions motivate follow-up observations to search for the presence of water vapor or oxygen on T1-c, and future observations of the outer planets in the TRAPPIST-1 system, which may possess substantial water.

Future model upgrades could include implementing a vertical atmospheric structure to enable more detailed escape rate calculations, additional atmospheric species, photochemistry, and surface sinks. As we move into a new era of observation, current work and future studies will assist in defining the plausible environments we could encounter, and statistical studies will be key to retrieving likely evolutionary histories.

\section{Acknowledgements}

We would like to thank the 2 anonymous peer reviewers whose suggestions greatly improved the quality of our manuscript. We also thank Miguel Morales and Bryna Hazelton for helpful advising on statistical methods, Josh Krissansen-Totton for providing data and discussion in our model comparison, David Catling for discussion on diffusion-limited escape, and Scott Anderson, Zeljko Ivezic, and the rest of UW's Qual Committee for thoughtful comments that helped improve the manuscript.

M.G., R.G, and J.B. acknowledge funding from the NSF Graduate Research Fellowship (DGE-2140004). R.B. acknowledges support from NASA grants 80NSSC20K0229 and 80NSSC23K0261. This work was supported in part by the  Virtual Planetary Laboratory Team, which is a member of the NASA Nexus for Exoplanet System Science, and funded via NASA Astrobiology Program Grant 80NSSC18K0829. 

%\begin{figure*}
%    \centering
%    \includegraphics[width=\textwidth]{Fullpage_percwaterlost.png}
%    \caption{Caption}
%    \label{fig:my_label}
%\end{figure*}

\appendix

%\vspace{-mm}
\section{AtmEsc} \label{appendix:atmesc}

The XUV environment determined by STELLAR serves as input to the atmospheric escape (AtmEsc) module of VPLanet \citep[][and further updated in this work]{LugerBarnes2015extreme,lincowski2018Trappist,Lincowski2022eratum,Barnes2020VPLanet}, here we describe our treatment of mass loss in the planetary atmosphere with AtmEsc. In mathematically describing hydrodynamic escape, we begin by defining the reference hydrogen flux ($F_{H}^{ref}$), which is the flux of hydrogen off the planet in the absence of any other species. It is prudent to start with this quantity as the reference hydrogen flux is used as foundation for every possible mass loss flux off the planet. In this work we are considering only pure water atmospheres initially, therefore the reference hydrogen flux is in the absence of oxygen. Note, the reference hydrogen flux is the true hydrogen flux for a planet in energy-limited escape and drag of a heavier species is shut off. Following \citet{LugerBarnes2015extreme} and \citet{schaefer2016predictions}, it is defined as:
\begin{equation}
    \large
    F_{H}^{ref} = \ddfrac{\epsilon_{XUV} F_{XUV} R_{p}}{4 G M_{p} \mu} \ .
\end{equation}
Here, $\epsilon_{XUV}$ is the XUV absorption efficiency, a fractional measure of the amount of incoming XUV flux that goes to driving escape; it is modelled following the treatment described in \citet{bolmont2017water}, which calculated $\epsilon_{XUV}$ with a set of 1D radiation-hydrodynamic mass-loss simulations\PeerReview{; we refer the reader to Figure 2 of \citet{bolmont2017water} for the range of adopted $\epsilon_{XUV}$ as a function of incoming XUV flux.} $F_{XUV}$ is the incoming XUV flux, $R_{p}$ is the radius of the planet, $M_{p}$ is the mass of the planet, and $\mu$ is the mass of one atomic mass unit (AMU).

After determining the reference hydrogen flux, the model will then determine if oxygen could be actively escaping via drag, the only oxygen sink directly modeled in this work. In energy-limited escape, drag occurs if the ``crossover mass'' ($m_{c}$) is greater than or equal to the mass of oxygen, and if the incoming XUV flux is greater than or equal to the critical drag XUV flux ($F_{XUV}^{Crit}$) \citep{Hunten1987massfracinescape}. Oxygen drag can also be shut off by the transition to diffusion-limited escape, but we will consider the aforementioned cases first. The crossover mass is the maximum particle mass that can be dragged by the escaping hydrogen flow; it is dependent on the reference hydrogen flux and, to fully describe the piecewise function of $m_{c}$, we first compare the reference hydrogen flux to an intermediary quantity, $x$ \citep{LugerBarnes2015extreme}:
\begin{equation}
    \large
    x = \left( \frac{m_{O}}{m_{H}} - 1 \right) \left( 1 - X_{O} \right) \left( \ddfrac{b_{diff} g \mu}{k_{b} T_{flow}} \right) \ .
\end{equation}
Here $m_{O}$ and $m_{H}$ are the masses of atomic oxygen and hydrogen, respectively, $X_{O}$ is the mixing ratio of oxygen, $g$ is the surface gravity, $T_{flow}$ is the temperature of the escaping flow, and $B_{diff}$ is the binary diffusion coefficient for atomic oxygen being dragged by a hydrogen outflow, which is given in [$m^{-1} s^{-1}$] by \citet{zahnle1986photochemistry} as:
\begin{equation}
    \large
    b_{diff} = 4.8 \times 10^{19} \ T_{flow}^{0.75} \ .
\end{equation}
Note, as our atmospheres are solely composed of oxygen and water, $1-X_{O}$ is equivalent to the mixing ratio of water ($X_{H_{2}O}$), which is a proxy for the mixing ratio of hydrogen. With our quantity $x$ in hand, we can define the crossover mass to be \citep{LugerBarnes2015extreme}:
\begin{equation}
%    \medium
    m_{c} =
    \begin{cases}
        \mu + \ddfrac{1}{1-X_{O}} \ \ddfrac{k_{b} T_{flow} F_{H}^{ref}}{b_{diff} g} & \text{if } F_{H}^{ref} < x \ , \\
        \mu \times \ddfrac{1+\left(\frac{X_{O}}{1-X_{O}}\right)\left(\frac{m_{O}}{m_{H}}\right)^{2}}{1+\left(\frac{X_{O}}{1-X_{O}}\right)\left(\frac{m_{O}}{m_{H}}\right)} + \ddfrac{k_{b} T_{flow} F_{H}^{ref}}{\left(1+X_{O}\left(\frac{m_{O}}{m_{H}}-1\right)\right) b_{diff} g} & \text{if } F_{H}^{ref} > x \ .
    \end{cases}
\end{equation}
The other parameter that determines oxygen drag is the minimum incoming XUV flux required for drag to occur, known as the critical drag XUV flux, and is defined by \citep{schaefer2016predictions}:
\begin{equation}
\label{eq:CritDrag}
    \large
    F_{XUV}^{Crit} = \ddfrac{4 b_{diff} \nu_{H}^{2}}{\epsilon_{XUV} k_{b} T_{flow} R_{p}} \left(\ddfrac{m_{O}}{m_{H}}-1\right) \left(1-X_{O}\right) \ .
\end{equation}

To summarize, oxygen drag will be shut off in the energy-limited regime if $F_{XUV} < F_{XUV}^{Crit}$ and/or $m_{c} < m_{O}$. If drag is occurring during energy-limited escape, the true hydrogen flux off the planet is less than the reference hydrogen flux and is given by:
\begin{equation}
\label{eq:FHdrag}
    \large
    F_{H}^{Drag} = F_{H}^{ref} \times \left(1 + \ddfrac{X_{O}}{1-X_{O}} \ \ddfrac{m_{O}}{m_{H}} \ \ddfrac{m_{c}-m_{O}}{m_{c}-m_{H}}\right)^{-1} \ .
\end{equation}
\PeerReview{While oxygen drag is occurring, we can define the flux of oxygen off the planet following \citet{LugerBarnes2015extreme}:
\begin{equation}
\label{eq:ofluxstart}
    \large
    F_{O} = \ddfrac{X_{O}}{1-X_{O}} F_{H} \left(\ddfrac{m_{c}-m_{O}}{m_{c}-m_{H}}\right) = \ddfrac{\eta}{2} F_{H} \ . 
\end{equation}
Where $\eta$ is defined as the ratio of oxygen escape to production:
\begin{equation}
\label{eq:etao}
    \large
    \eta = \ddfrac{2 X_{O}}{1-X_{O}} \ \ddfrac{m_{c} - m_{O}}{m_{c} - m_{H}} \ .
\end{equation}}

%When drag is occurring, the net flux of oxygen in and out of the atmosphere is given by:
%\begin{equation}
%\label{eq:oxyflux}
%    \large
%    F_{O} = \ddfrac{X_{O}}{1-X_{O}} F_{H} \left(\ddfrac{m_{c} - m_{O}}{m_{c} - m_{H}}\right)
%\end{equation}
%Unlike the implementation of AtmEsc in \citet{LugerBarnes2015extreme}, if there is a built-up oxygen reservoir we allow oxygen to be lost through drag faster than it is photolytically produced at a given timestep; inordinate oxygen drag is shown to be realistic in several previous studies \citep[e.g.,][]{bolmont2017water,Johnstone2020hydrodynamicesc}. 

Drag will also be halted in the diffusion-limited regime because it is defined when the background heavier gas is static, meaning the escape flux of hydrogen has been throttled sufficiently to prevent drag of a heavier species \citep{hunten1973escape}. Following \citet{schaefer2016predictions}, the switch to diffusion-limited escape occurs if the mixing ratio of free oxygen (i.e., oxygen not in water molecules) in the atmosphere is greater than or equal to the mixing ratio of water ($X_{O} \ge X_{H_{2}O}$). In this case, the true hydrogen flux off the planet again is less than the reference hydrogen flux and is given by \citep{LugerBarnes2015extreme}:
\begin{equation}
    \large
    F_{H}^{Diff} = b_{Diff} g \mu \ddfrac{\frac{m_{O}}{m_{H}}-1}{k_{b} T_{flow} \left(1+\frac{X_{O}}{1-X_{O}}\right)} \ .
\end{equation}

Now that we have defined all possible hydrogen flows off the planet, we can define the total water loss over time assuming the planetary radius is the effective XUV radius \citep{LugerBarnes2015extreme}:
\begin{equation}
    \large
    \dot M = F_{H} 4 \mu \pi R_{p}^{2} \ ,
\end{equation}
%where $F_{H}$ refers to whatever the true hydrogen flux is off the planet at a given time. To calculate the oxygen flux in the atmosphere at a given time, we first define a parameter, $\eta$, that is defined as the ratio of oxygen escape to production; following \citet{LugerBarnes2015extreme}:

%Finally, the net rate of oxygen flux in the atmosphere is given by:
\PeerReview{This further allows us to rewrite the oxygen flux off the planet during drag (equation \ref{eq:ofluxstart}) in terms of the water loss rate:}
\begin{equation}
    \large
    F_{O} = \ddfrac{8 - 8 \eta}{1 + 8 \eta} \ \dot M \ .
\end{equation}

\PeerReview{\subsection{Model Validation} \label{subsec:ModelCompare}}

\begin{figure}
    \centering
    \includegraphics[width=\textwidth]{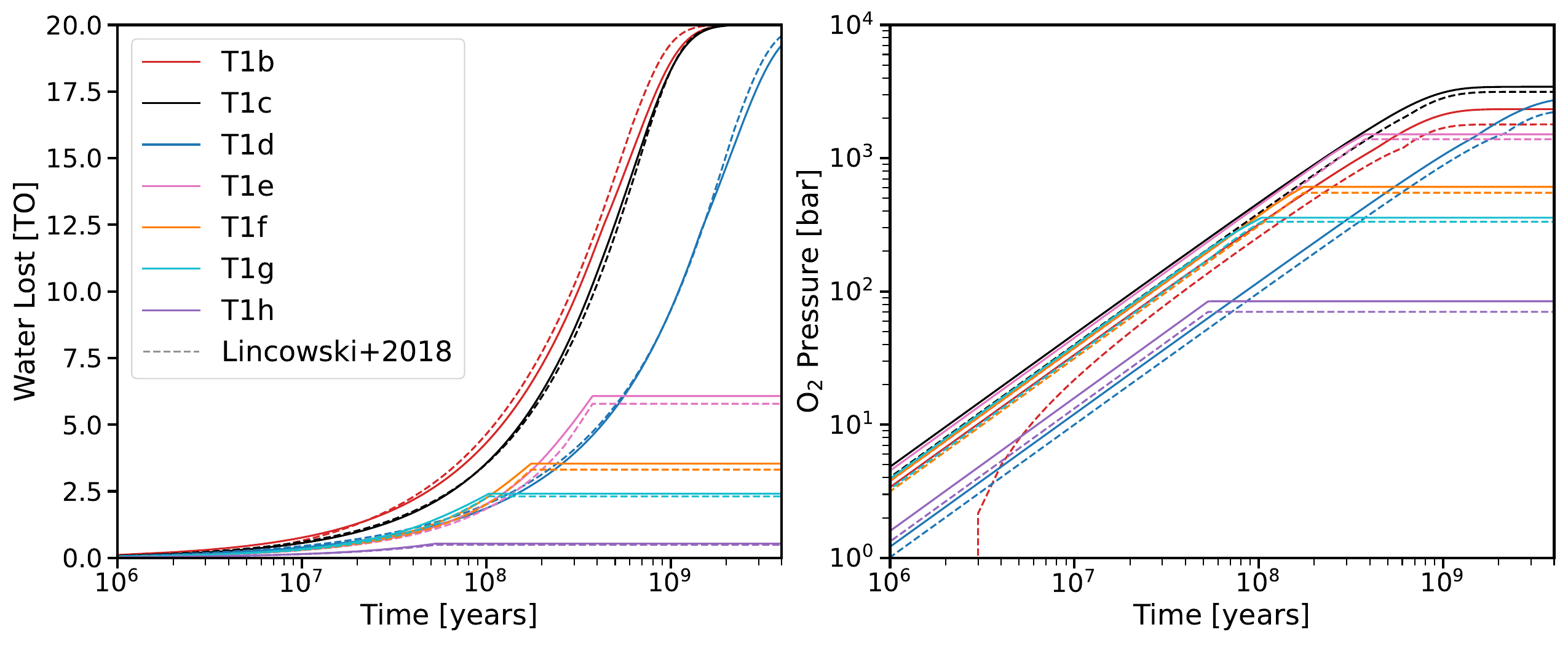}
    \caption{A comparison to the escape rates of the TRAPPIST-1 planets calculated in \citet{lincowski2018Trappist,Lincowski2022eratum} using constraints on the system from \citet{Grimm2018trappist}. \textbf{Left Plot:} The water lost in TO over time in years beginning with 20 TO initially. \textbf{Right Plot:} The oxygen produced in bars over time in years. Solid lines denote the rates calculated using the new version of AtmEsc from this work, and dashed lines denote the rates calculated using the version of AtmEsc from \citet{lincowski2018Trappist,Lincowski2022eratum}.}
    \label{fig:linccompare}
\end{figure}

To show how the modifications to AtmEsc in this work have changed the outputs of the model from the original version \citep{LugerBarnes2015extreme,lincowski2018Trappist,Lincowski2022eratum,Barnes2020VPLanet}, we present a comparison of one simulation from the original version of AtmEsc to the version modified in this work. Figure \ref{fig:linccompare} shows an escape rate calculation for a single simulation of the TRAPPIST-1 planets with an initial water content of 20 TO from \citet{lincowski2018Trappist,Lincowski2022eratum}, compared to the same calculation done with our model upgrades using the same input parameters. Note, as \citet{lincowski2018Trappist,Lincowski2022eratum} used inputs for the TRAPPIST-1 system defined by \citet{Grimm2018trappist}, this comparison is meant to show the differences in the models and does not use the updated constraints on the system we adopt in the rest of our work. Additionally, we adopt the same XUV model in this calculation as \citet{lincowski2018Trappist,Lincowski2022eratum}.

There are only minor differences between the two versions of AtmEsc in our comparison. In both models T1-b and c become desiccated, and T1-d is nearly desiccated, losing $\sim$0.14 TO more water in \citet{lincowski2018Trappist,Lincowski2022eratum}. The three interior planets all produce $\sim$10 -- 30\% more oxygen in this work; this is because \citet{lincowski2018Trappist,Lincowski2022eratum} predicts faster desiccation on these planets, meaning oxygen drag is more efficient during periods of greater water loss preventing build-up. The outer planets all lose more water in this work, typically by $\leq$0.2 TO, and as a result build up more oxygen, typically by $\leq$100 bars. 
%Overall, we have shown that modifications to AtmEsc in this work produce results commensurate with those reported by older versions of the model.

\section{Outlier Removal} \label{appendix:outliers}

\begin{figure}
    \centering
    \includegraphics[width=0.7\textwidth]{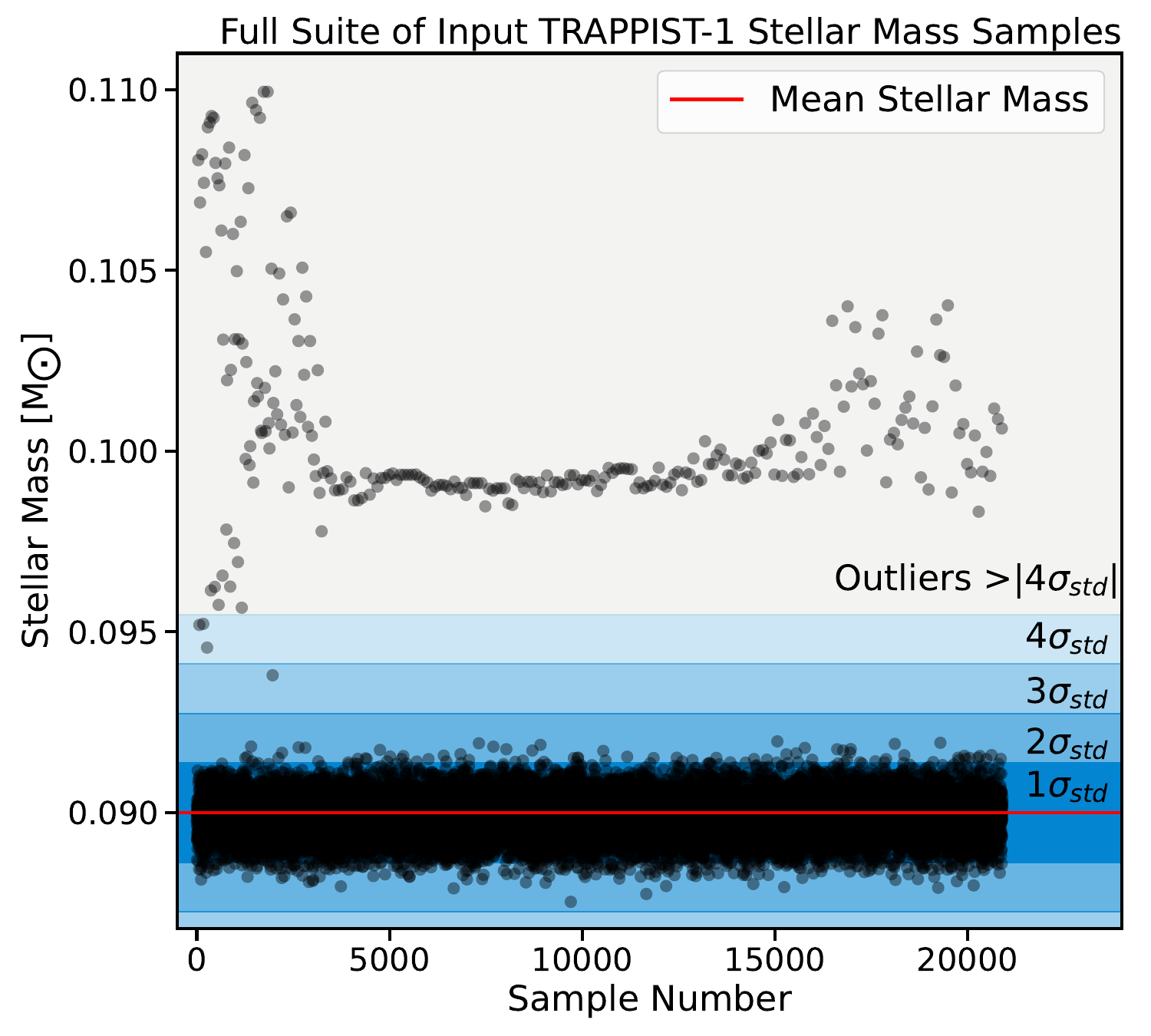}
    \caption{All 20,900 stellar mass samples available for simulations to draw from \citep[from][]{birky2021improved}. The y-axis shows the stellar mass in M$_{\bigodot}$ of a given sample, which are arbitrarily assigned an ID on the x-axis. The red solid line denotes the mean stellar mass across all samples, while the dark blue and subsequent lighter blue regions denote 1, 2, 3, and 4 standard deviations from the mean, as labeled in the plot. The light grey shaded region denotes samples greater than 4 standard deviations from the mean, indicating that samples in this region are classified as outliers, such that simulations using stellar masses above 4 standard deviations are removed from the final data set.}
    \label{fig:outliers}
\end{figure}

Within our full dataset across all initial water contents, a small subset of simulations were identified as outliers originating from the stellar mass samples. As discussed in Section \ref{subsec:Input}, we sample stellar and planetary inputs from \citet{birky2021improved} and \citet{agol2021refining}, respectively, creating 20,900 input strings a simulation could draw from. Figure \ref{fig:outliers} shows the stellar mass samples plotted on the y-axis, with an arbitrarily assigned ID number on the x-axis. The majority of samples fall within 2 standard deviations from the mean, but a small group of samples lie at stellar masses greater than 4 standard deviations from the mean, which we adopt as our outlier cut-off. Outliers account for 259/20900, or 1.24\%, of all samples. Simulations using a stellar mass above this cut-off are removed from the final dataset.

These outliers originate from the process used in \citet{birky2021improved} to fit the stellar mass of TRAPPIST-1 with the most recent observational data. They used a Gaussian process (GP) surrogate model to estimate the probability of an input parameter, including stellar mass, given the observed stellar properties. Sampling of the GP surrogate posterior was performed using the emcee package \citep{ForemanMackey2013emcee}. We found that these outliers originated from the randomized initial state of the MCMC walkers, suggesting that number of burn-in samples reported in the posterior samples by \citet{birky2021improved} was insufficient. The burn-in period was taken to be 2x the average autocorrelation time, which appeared to not reflect the converged state for a subsample of the MCMC walkers. Thus, these outlying stellar mass samples are likely a product of randomized initial walker states, and as they account for 1.24\% of all samples, we conclude their removal is justified.

\section{Predicting Initial Water Content from a Final State} \label{appendix:initialpredictionMethods}

In Section \ref{subsec:InitialWaterPredict} we showed the most likely initial water content our model would predict to reproduce a known present-day predicted or observed water content on the planet. Here, we detail the generalized method used to achieve this result.

To start, let's assume we have constraints on a planet's present-day water content from either observation or simulation, and models that map initial water content to final water content. For each initial water content in our sampling grid, there are a sufficient amount of simulations (typically $\sim$3500 -- 5500 in our work) to encompass the uncertainty in the planet and host star’s physical parameters, which in turn define the uncertainty in the final water content at a particular initial content. The question we then pose here is how to determine which initial water content on our sampling grid best produces the final state from our present-day constraints.

The first step is to model the planet's present-day water constraints with a probability density function (PDF) to assign probability weighting to final water contents based on the defined uncertainty in the constraints. In this work, we tested several common statistical functions and identified a gamma profile to best fit the present-day constraint we adopt. Gamma profiles worked well since the water mass fractions are restricted to be greater than or equal to 0 and, consequently, show asymmetric tails, but any continuous probability distribution may be used here in theory. With a gamma profile, our present-day water PDFs are described by:
\begin{equation}
\label{eq:pdf}
    \large
    f_{i}(w,a) = \ddfrac{w^{a-1} e^{-w}}{\Gamma(a)} \ ,
\end{equation}
where $i$ denotes which planet the PDF is describing (T1-b, c, d, e, f, g, or h), $w$ is the present-day water content, $a$ is the shape parameter, $f_{i}(w,a)$ is the relative likelihood of present-day water content $w$ on planet $i$, and $\Gamma(a)$ is the gamma function evaluated at a particular shape parameter.

With the PDF describing the present-day water constraints, the next step is to determine the value of the PDF at the final water content predicted by each individual simulation across a particular initial water content. We then sum the PDF value across the individual simulations of a given initial water content and normalize for the total number of simulations in that suite to obtain a probability weighting for that particular initial water content. To normalize the probability weightings per initial water content, we divide by the number of simulations in the PDF value summation:
\begin{equation}
\label{eq:normalizedlikelihood}
    \large
    \text{Normalized Likelihood} = \ddfrac{\sum_{z=1}^{N} f_{i}(w_{z}, a)}{N} \ , \ \text{per initial water content},
\end{equation}
where $f_{i}(w_{z}, a)$ is the PDF for the planet being considered (Equation \ref{eq:pdf}) evaluated at the final water content of a particular simulation, $w_{z}$, and $N$ is the total number of simulations at that initial water content.

The initial water content with the highest probability weighting is the one that best produced the assumed present-day water constraints. If the initial water content grid is sampled finely enough, we can interpolate across our distribution of probability weightings to determine where the true most likely initial water content would be, and what the confidence intervals are on that prediction. This next step can only be taken if the grid of initial water contents with simulation data is fine enough that the distribution of probability weightings appears to be a smooth, continuous curve; in some cases this may require running more data once the initial peak probability weighting has been identified. With sufficient data, this interpolation can be accomplished with a Riemann sum. To perform the Riemann sum, we take the curve of probability weightings versus initial water content and create area bins based on the data points, then find at what initial water content is the area under the curve 50\% of the total area -- the 50$^{th}$ percentile, or median value (i.e., the initial water content most likely to produce the present-day constraints). Then finding the bounds of 68\% and 95\% area under the curve, centered at the median value, will yield the 1$\sigma$ and 2$\sigma$ confidence intervals.

In our work, we created bins between adjacent data points with linear lines as the top of the bins. Thus, the area in a particular bin between two initial water contents, $IW1$ and $IW2$, is given by:
\begin{equation}
    \large
    A_{bin} = \int_{IW1}^{IW2} \left(sx + b\right) dx = \left(\ddfrac{s}{2} x^{2} + bx \right) \bigg|_{IW1}^{IW2} = \ddfrac{s}{2} \left( IW2^{2} - IW1^{2} \right) + b \left( IW2 - IW1 \right) \ ,
\end{equation}
where $s$ is the slope of the linear fit between the two data points and $b$ is the y-intercept. This integrand is the same equation used to solve for a true value or confidence interval bound that lies between two data points. In fitting for the PDF this work, we used SciPy's stats module \citep{SciPy2020}, and NumPy's polyfit routine \citep{NumPy2020} for the linear fits. 

%This method is implemented in the current study in Section \ref{subsec:InitialWaterPredict}, and the implication of those results, broader application, and importance of this method is discussed in Section \ref{subsec:PredictingInitialDiscuss}.
                             

\bibliography{main}{}
\bibliographystyle{aasjournal}



\end{document}